\documentclass[iop,numberedappendix]{emulateapj}
\usepackage{apjfonts}
\newcommand{\be}{\begin{equation}}
\newcommand{\ee}{\end{equation}}
\newcommand{\Msun}{M_{\odot}}
\def\kms{\ {\rm km\, s}^{-1}}
\def\teff{T_{\rm eff}}

\newcommand\ionn[2]{#1$\;${\scshape{#2}}}

\shortauthors{CONROY, GRAVES \& VAN DOKKUM}

\shorttitle{Early-Type Galaxy Archeology}

\begin{document}

\title{Early-Type Galaxy Archeology: Ages, Abundance Ratios, and
  Effective Temperatures from Full-Spectrum Fitting}

\author{Charlie Conroy\altaffilmark{1},
  Genevieve J. Graves\altaffilmark{2},
  and Pieter G. van Dokkum \altaffilmark{3}}

\altaffiltext{1}{Department of Astronomy \& Astrophysics, University
  of California, Santa Cruz, CA, USA}
\altaffiltext{2}{Department of Astrophysical Sciences, Princeton
  University, Princeton, NJ, USA}
\altaffiltext{3}{Department of Astrophysical
  Sciences, Yale University, New Haven, CT, USA}
\slugcomment{Submitted to ApJ}

\begin{abstract}

  The stellar populations of galaxies hold vital clues to their
  formation histories.  In this paper we present results based on
  modeling stacked spectra of early-type galaxies drawn from the Sloan
  Digital Sky Survey (SDSS) as a function of velocity dispersion,
  $\sigma$, from $90\kms$ to $300\kms$.  The spectra are of extremely
  high quality, with typical signal-to-noise ratio, S/N, of
  1000\,\AA$^{-1}$, and a wavelength coverage of 4000\AA$-$8800\AA.
  Our population synthesis model includes variation in 16 elements
  from C to Ba, a two-component star formation history, the shift in
  effective temperature, $\Delta\,\teff$, of the stars with respect to
  a solar metallicity isochrone, and the stellar initial mass function
  (IMF), amongst other parameters.  In our approach we fit the full
  optical spectra rather than a select number of spectral indices and
  are able to, for the first time, measure the abundances of the
  elements V, Cr, Mn, Co, and Ni from the integrated light of distant
  galaxies.  Our main results are as follows: 1) light-weighted
  stellar ages range from $6-12$ Gyr from low to high $\sigma$; 2)
  [Fe/H] varies by less than 0.1 dex across the entire sample; 3) Mg
  closely tracks O, and both increase from $\approx0.0$ at low
  $\sigma$ to $\sim0.25$ at high $\sigma$; Si and Ti show a shallower
  rise with $\sigma$, and Ca tracks Fe rather than O; 4) the iron peak
  elements V, Cr, Mn, and Ni track Fe, while Co tracks O, suggesting
  that Co forms primarily in massive stars; 5) C and N track O over
  the full sample and [C/Fe] and [N/Fe] exceed $0.2$ at high $\sigma$
  ; and 6) the variation in $\Delta\,\teff$ with total metallicity
  closely follows theoretical predictions based on stellar evolution
  theory.  This last result is significant because it implies that we
  are robustly solving not only for the detailed abundance patterns
  but also the detailed temperature distributions (i.e., isochrones)
  of the stars in these galaxies.  A variety of tests reveal that the
  systematic uncertainties in our measurements are probably 0.05 dex
  or less.  Our derived [Mg/Fe] and [O/Fe] abundance ratios are
  $0.05-0.1$ dex lower than most previous determinations.  Under the
  conventional interpretation that the variation in these ratios is
  due to star formation timescale variations, our results suggest
  longer star formation timescales for massive early-type galaxies
  than previous studies.  Detailed chemical evolution models are
  necessary in order to translate the abundance ratio distributions of
  these galaxies into constraints on their formation histories.
  Alternatively, these data may provide useful constraints on the
  nucleosynthetic pathways for elements whose production is not well
  understood.

\end{abstract}



\section{Introduction}
\label{s:intro}

We now have a working theory for the evolution of large scale
structure in the universe thanks to the largely successful paradigm of
a universe dominated by cold dark matter.  In this paradigm
cosmological structures form ``bottom-up'' --- small structures
collapse early and the massive clusters assemble late.  One of the
more interesting puzzles in this paradigm is the realization that the
most massive galaxies in our universe seem to have formed their stars
earliest, with lower-mass galaxies forming their stars later
\citep[e.g.,][]{Trager98, Thomas05, Jimenez07}.  This phenomenon goes
by many names, including ``downsizing'' and ``anti-hierarchical
growth''.  No {\it ab initio} model for galaxy formation is able to
reproduce these basic features of the observed galaxy population.  The
apparent tension between the bottom-up nature of cold dark matter and
the top-down star formation histories of galaxies can be reconciled by
recognizing that the {\it formation} time of the stars need not be
related to the {\it assembly} time of the stars into their present day
galaxies.

From an empirical point of view, the formation and evolution of
galaxies can be probed via two general techniques.  The first is
through lookback studies where one observes, statistically, the
progenitors of present day galaxies at progressively higher redshifts.
The second is through studying the present day properties of galaxies,
including their stellar populations, structure, and kinematics, in
order to learn about their past evolution.  This latter technique is
often referred to as the archeological approach, and is the subject of
this paper.

\begin{figure*}[!t]
\center
\resizebox{7in}{!}{\includegraphics{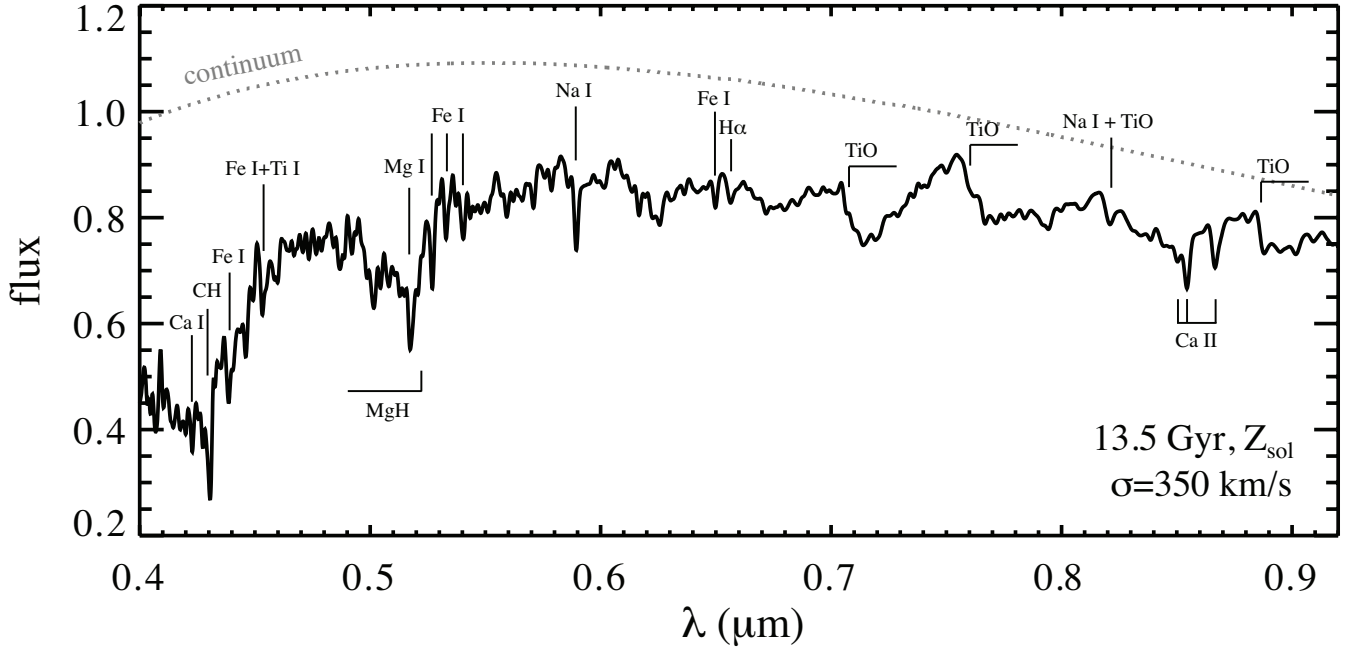}}
\caption{Model spectrum for an age of 13 Gyr and solar metallicity.
  The spectrum has been smoothed with a velocity dispersion of
  $\sigma=350\kms$, equal to the smoothing applied to the early-type
  galaxy data analyzed in this paper.  Strong features are labeled.
  Also included is the location of the true stellar continuum, which
  is the spectrum that would be observed in the absence of all line
  opacity.  In this figure the model spectrum is computed entirely
  from synthetic stellar spectra, whereas for the main analysis the
  synthetic spectra are only used differentially.}
\label{fig:spec}
\end{figure*}

\begin{figure*}[!t]
\center
\resizebox{7in}{!}{\includegraphics{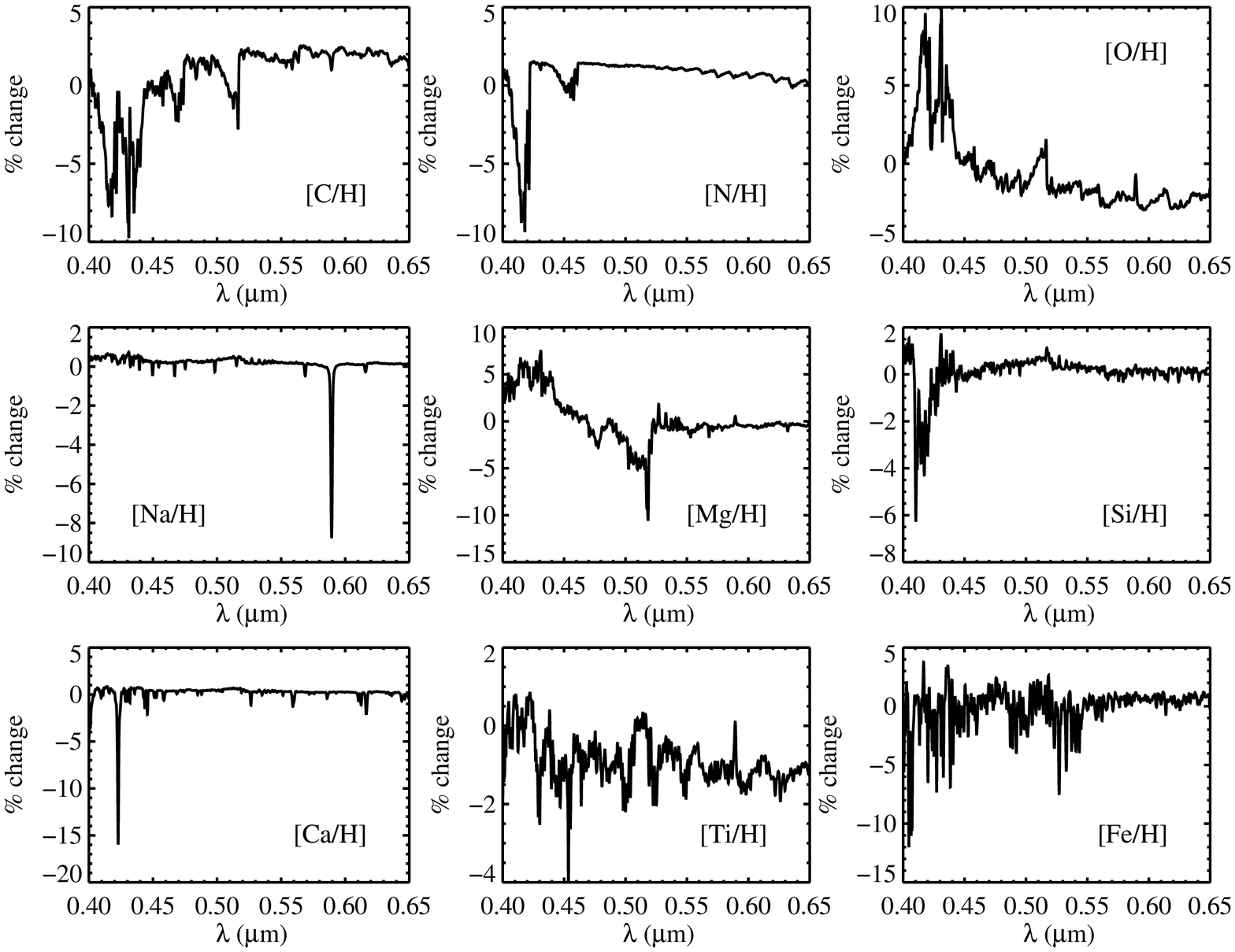}}
\caption{Relative response of the blue spectral region to an increase
  in the abundance of nine different elements.  Abundances have been
  increased by 0.3 dex for all elements except for C, which is
  increased by 0.15 dex.  The changes are with respect to a solar
  metallicity model with a MW IMF and an age of 13 Gyr.  The models
  have been smoothed to a velocity dispersion of $\sigma=150\kms$.
  Notice the different y-axis range in each panel.  These are
  noiseless spectra, and so every feature visible in the figure is
  real.  It is evident from both this figure and Figures \ref{fig:red}
  and \ref{fig:fepeak} that essentially every spectral region is
  influenced by one or more elements. }
\label{fig:blue}
\end{figure*}

It has long been appreciated that optical through near-infrared (NIR)
spectra of old stellar systems are rich in absorption features that
provide clues to the ages, metallicities, and abundance patterns of
the stars \citep[for a review, see][]{Conroy13b}. Since the 1980s the
standard analysis technique has been to measure and model the strength
of stellar absorption features through the Lick/IDS index system
\citep{Burstein84, Worthey94b}.  Authors typically model a handfull of
these indices in order to estimate various stellar population
properties.

\begin{figure*}[!t]
\center
\resizebox{7in}{!}{\includegraphics{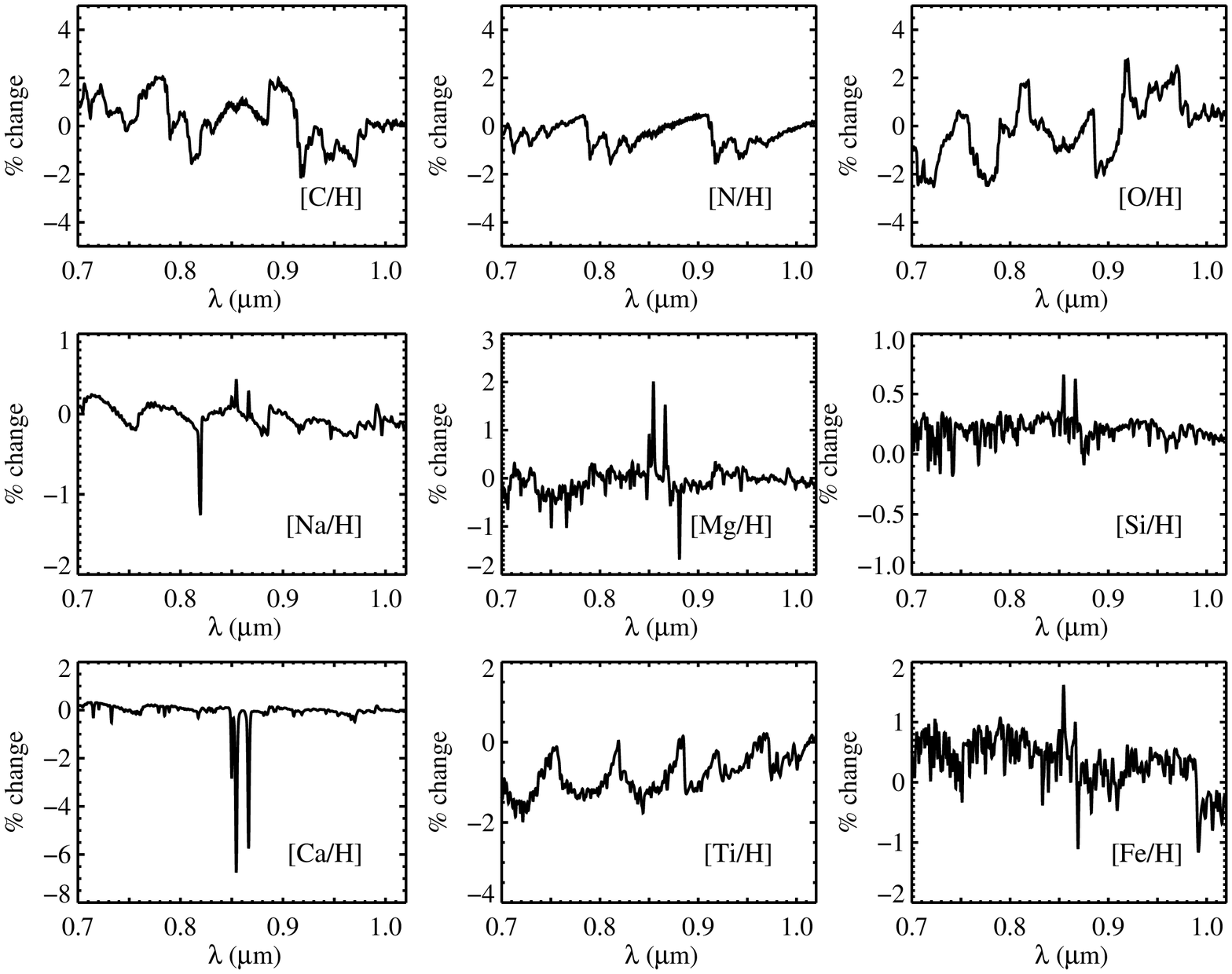}}
\caption{Same as Figure \ref{fig:red}, now showing the red spectral
  region.  Abundances have been increased by 0.3 dex for all elements
  except for C, which is increased by 0.15 dex}
\label{fig:red}
\end{figure*}

Analysis of the Lick indices has revealed that early-type galaxies
are enhanced in the $\alpha$ element Mg compared to the abundance
patterns of stars in the Galactic disk \citep{Worthey94}.  The
[$\alpha$/Fe] ratio has long been known to be sensitive to several
parameters, including the timescale of star formation, the slope of
the initial mass function (IMF) at $>1\Msun$, the delay time
distribution of Type Ia supernovae (SNe), and the preferential loss of
metals via winds \citep[e.g.,][]{Tinsley79, Thomas99}.  The
sensitivity of [$\alpha$/Fe] to all of these processes significantly
complicates the interpretation of this ratio.  Nonetheless, it has
become standard practice to assume that the [$\alpha$/Fe] ratio is
telling us something about the timescale of star formation.  By
comparing to simple closed-box chemical evolution models,
\citet{Thomas05} concluded that the [$\alpha$/Fe] ratios of the most
massive galaxies in their sample, with $\sigma\approx300\kms$, implied
star formation timescales of only $\sim200$ Myr.  These short inferred
star formation timescales, if correct, would have profound
implications for our understanding of the formation of the most
massive galaxies in the universe.

The analysis of spectral indices is beginning to give way to the
modeling of the full optical-NIR spectra of stellar systems.  This
approach has great potential, and is now frequently used to measure
star formation histories and metallicities non-parametrically
\citep{Heavens00, CidFernandes05, Ocvirk06, Tojeiro09}.  Recently,
such full spectrum fitting models have been extended to include
variation in the elemental abundance patterns \citep{Walcher09,
  Conroy12a}, paving the way to robustly measure many parameters from
integrated light spectra.

This paper is part of an ongoing series focused on modeling very high
quality optical-NIR spectra of early-type galaxies with full-spectrum
fitting stellar population synthesis (SPS) techniques.  In
\citet[][CvD12]{Conroy12a}, we presented the model, and in
\citet{Conroy12b} we used that model to measure the low-mass IMF from
the spectra of 34 nearby early-type galaxies and the nuclear bulge of
M31 \citep[based on data presented in][]{vanDokkum12}.  In
\citet{Conroy13a} we presented results on the neutron-capture elements
Sr and Ba based on our local galaxy sample and on the stacked spectra
that are the focus of this work.  In the present paper we measure the
ages, detailed abundance patterns, and effective temperatures of the
stars in early-type galaxies drawn from the SDSS.  In a subsequent
paper we will discuss the low-mass IMFs inferred for these galaxies.

The rest of this paper is organized as follows.  Section \ref{s:model}
provides an overview of the SPS model and our fitting technique,
Section \ref{s:data} describes the data, and Section \ref{s:calib}
presents a test of the model by fitting to spectra of metal-rich star
clusters.  Our main results are presented in Section \ref{s:res}, and
in Section \ref{s:comp} we compare our derived properties to results
based on other modeling techniques.  A discussion and summary of our
results is provided in Sections \ref{s:disc} and \ref{s:sum}.

\section{Model \& Fitting Technique}
\label{s:model}

\begin{figure*}[!t]
\center
\resizebox{7in}{!}{\includegraphics{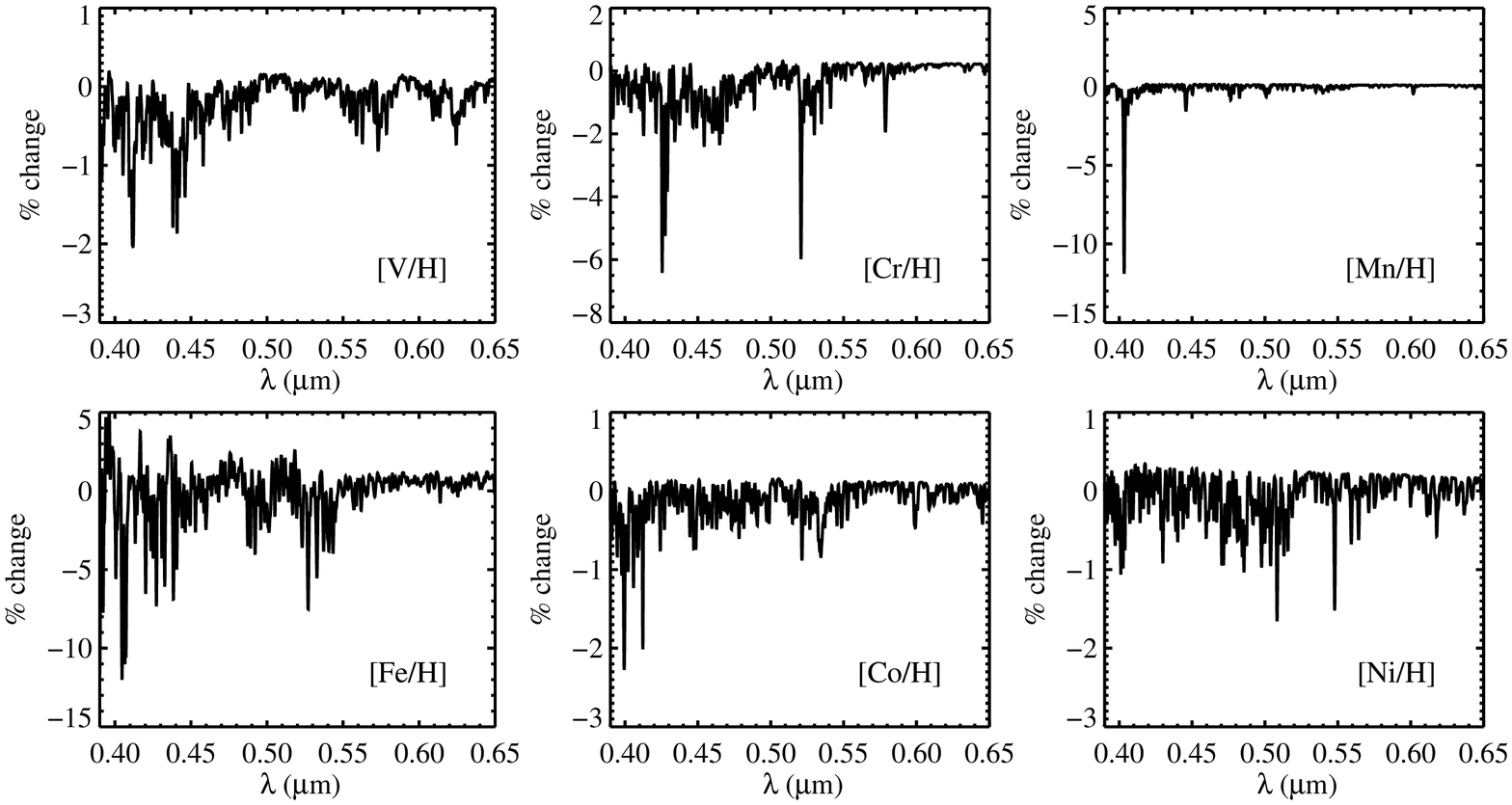}}
\caption{Same as Figure \ref{fig:blue}, now showing the elements
  encompassing the iron peak (V, Cr, Mn, Fe, Co, Ni).  The change in
  the model spectra have been computed for a 0.3 dex increase in the
  abundance of each element.}
\label{fig:fepeak}
\end{figure*}

\begin{figure}[!t]
\center
\resizebox{3.5in}{!}{\includegraphics{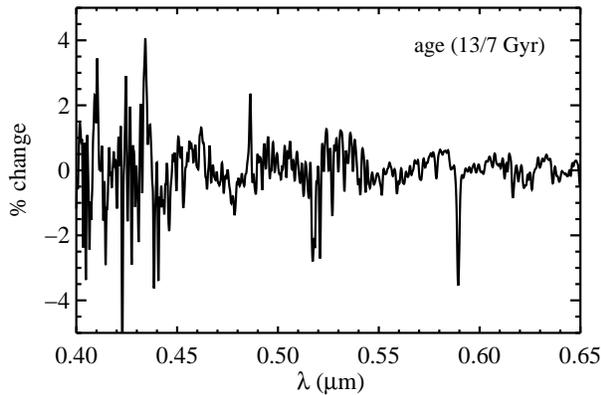}}
\caption{Relative response of the blue spectral region to an age
  variation from 7 to 13 Gyr.  The model has solar metallicity and has
  been smoothed to a velocity dispersion of $\sigma=150\kms$}
\label{fig:age}
\end{figure}

\begin{figure}[!t]
\center
\resizebox{3.5in}{!}{\includegraphics{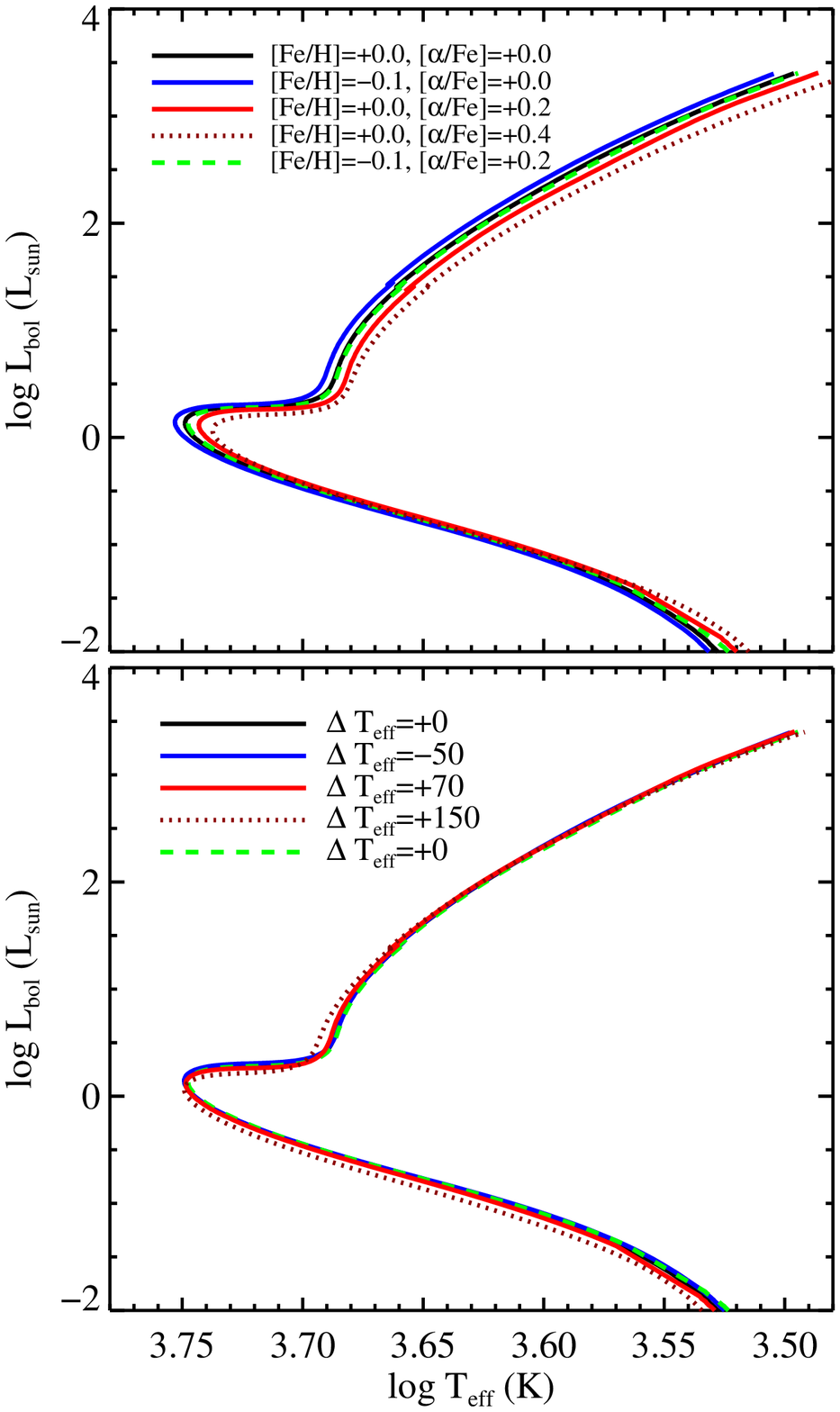}}
\caption{Isochrones at 13.4 Gyr from the Dartmouth Stellar Evolution
  Database (DSEP) for a variety of chemical compositions.  {\it Top
    panel:} Isochrones for a range in compositions approximately
  spanned by the early-type galaxies in our sample.  The case with
  [Fe/H]$=0.0$ and [$\alpha$/Fe]$=+0.4$ is extreme, as no galaxies in
  the sample have such large $\alpha$ enhancements.  Notice that the
  isochrone with [Fe/H]$=-0.1$ and [$\alpha$/Fe]$=0.2$ is almost
  identical to the solar metallicity isochrone. {\it Bottom panel:}
  Same isochrones as in the top panel, with a {\it constant} shift in
  $\teff$ applied to the entire isochrone (the shift applied is shown
  in the legend).  For the modest range in chemical compositions
  probed by our sample, the expected effective temperatures of the
  stars can therefore be well-approximated by a solar metallicity
  model with a single additional free parameter describing the offset
  in $\teff$.}
\label{fig:iso}
\end{figure}

The SPS model used herein was developed in \citet[][CvD12]{Conroy12a},
to which we refer the interested reader for details.  The model
employs standard SPS techniques, including combining libraries of
isochrones and stellar spectra.  We use isochrones from three separate
groups, all of which are solar metallicity with solar-scaled abundance
patterns.

Empirical spectra form the base of the model.  We make use of two
separate libraries, the MILES library, which covers the wavelength
range $0.35\mu m<\lambda<0.74\mu m$ \citep{Sanchez-Blazquez06}, and
the IRTF library of cool stars, which covers the wavelength range
$0.81\mu m<\lambda<2.4\mu m$ \citep{Cushing05, Rayner09}.  The
empirical stars are of approximately solar metallicity and have solar
abundance patterns.  We have computed a large grid of synthetic
stellar atmospheres and spectra in order to model the relative change
in the spectrum of a star due to a change in the abundance of a single
element.  We chose 20 positions along a 13 Gyr isochrone to compute
synthetic spectra for each abundance pattern.  The model atmospheres
and spectra were computed with the ATLAS12 model atmosphere and
spectrum synthesis package \citep{Kurucz70, Kurucz93}, ported to Linux
by \citet{Sbordone04}.  The line list was provided by
R. Kurucz\footnote{\texttt{kurucz.harvard.edu/}}, including linelists
for TiO and H$_2$O, amongst other molecules.  The synthetic stellar
spectra were grafted onto the empirical stellar spectra in order to
create models with arbitrary abundance patterns.  We emphasize that
the synthetic spectra are only used differentially.  The model allows
for arbitrary variation in the IMF and spans ages from $3-13.5$ Gyr.
Figure \ref{fig:spec} shows the model spectrum for a 13 Gyr, solar
metallicity population, smoothed to a velocity dispersion of
$\sigma=350\kms$.

There have been several updates to the CvD12 model.  Most importantly,
the synthetic atmospheres and spectra have been recomputed with an
updated line list kindly provided by R. Kurucz.  This line list
includes many more iron lines and updated wavelengths and hyperfine
splitting of many lines.  The model now contains variation in many
more elements, described below. We have also supplemented the MILES
empirical library with stacked SDSS M dwarf spectra from
\citet{Bochanski07}.  This turns out to be important because the MILES
library contains only a few M dwarf spectra, and none cooler than M6V.
The SDSS M dwarf template spectra cover the entire M dwarf spectral
sequence.  The final update since CvD12 is the inclusion of emission
lines in the fitting (rather than masking regions of potential
emission line contamination, as done previously).  Currently we
include 13 emission lines, including individual components of
[\ionn{O}{ii}]$\lambda$3727, [\ionn{Ne}{iii}]$\lambda$3868, H$\delta$,
H$\gamma$, H$\beta$, [\ionn{O}{iii}]$\lambda$5007,
[\ionn{N}{i}]$\lambda$5201, [\ionn{N}{ii}]$\lambda$6547,6583,
H$\alpha$, and [\ionn{S}{ii}]$\lambda$6716,6731.  We include separate
parameters for the strengths of each of these lines (i.e., we do not
enforce particular line ratios, even amongst doublets with known,
constant line ratios).  The intrinsic line shape is assumed to be a
gaussian with a single width for all lines.  This width is also
included in the fit.

The effect of varying elemental abundances on the model spectra are
shown in Figures \ref{fig:blue}, \ref{fig:red}, and \ref{fig:fepeak}.
These figures show the response of a 13 Gyr, solar metallicity model
to an increase in the abundance of a single element.  We emphasize
that the model atmospheres and spectra are always treated
self-consistently (i.e., the elemental concentration is increased both
in the computation of the model atmosphere and of the model spectra).
In many cases the effect on the spectrum from a change in an element
is straightforward, e.g., increasing [Mg/H] results in stronger
\ionn{Mg}{i} lines, but other effects are more subtle.  For example,
because Mg is a major electron donor, increasing [Mg/H] affects the
ionization equilibrium and therefore affects many species of other
elements.  For example, it is for this reason that [Mg/H] affects the
strength of the \ionn{Ca}{ii} lines at $\sim0.86\mu m$.

The effect of varying age from 7 to 13 Gyr is shown in Figure
\ref{fig:age}.  A variety of temperature-sensitive features are
evident, including the hydrogen balmer lines, NaD, and the
\ionn{Mg}{i} lines.  Most importantly, comparison of Figure
\ref{fig:age} to Figures \ref{fig:blue}-\ref{fig:fepeak} shows that
when one considers information in the full spectrum, age and
metallicity effects are readily separable.

We follow \citet{Conroy12b} in fitting the model to data.  In its
present form the model contains 40 free parameters, including the
redshift and velocity dispersion, a two-part power-law IMF, two
population ages (the age of the dominant population and the mass
fraction of a 3 Gyr population), four nuisance parameters, 13 emission
line strengths, the velocity broadening of the emission lines, and the
abundances of C, N, Na, Mg, Si, Ca, Ti, V, Cr, Mn, Fe, Co, Ni, Sr, and
Ba, and O, Ne, S (the latter three are varied in lock-step).  These
parameters are fit to the data via a Markov Chain Monte Carlo fitting
technique.  The data and models are split into four wavelength
intervals (described below) and, within each interval the spectra are
normalized by a high-order polynomial \citep[with degree $n$ where
$n\equiv (\lambda_{\rm max}-\lambda_{\rm min})/100$\AA; see][for
details]{Conroy12b}.  For the purposes of the present article, the
emission line strengths are treated as additional nuisance parameters,
and in many cases their strengths are constrained to be very low,
which is not surprising given the sample definition described in the
next section.  Unless specified otherwise, ages quoted in this paper
are light-weighted stellar ages computed from the two age components.

One of the nuisance parameters included in the model is a shift in
effective temperature, $\Delta\,\teff$, applied to all stars used in
the synthesis.  As described in CvD12, this parameter is meant to
capture changes in the isochrones due to changes in abundance patterns
of the model stars.  The motivation for this shift is shown in Figure
\ref{fig:iso}.  In the top panel we show isochrones obtained from the
Dartmouth Stellar Evolution Database \citep[DSEP;][]{Dotter08b} for a
range in [Fe/H] and [$\alpha$/Fe].  In the bottom panel we have
shifted each isochrone by a constant amount in $\teff$ until a match
was achieved with the solar metallicity case.  A constant shift does
an excellent job of accounting for modest variation in abundance
patterns from the solar metallicity model \citep[see
also][]{Salaris93}.  We caution that simple shifts in $\teff$ work
less well at low metallicity \citep{VandenBerg12}.  Moreover, it is
interesting that the isochrone with [Fe/H]$=-0.1$ and
[$\alpha$/Fe]$=0.2$ is almost identical to the solar metallicity
isochrone.  Thus, over the metallicity range of interest in this work,
a simple shift in $\teff$ is probably sufficient to account for
elemental abundance effects on the isochrones.  Because the abundances
of many elements are varied simultaneously in the model, we use the
free parameter $\Delta\,\teff$ to encompass their combined effect,
rather than attempting to compute self-consistent isochrones from
first principles.  However, we caution that a constant shift in
$\teff$ is unlikely to capture the full variation in isochrone
morphology with elemental abundance.  For example, a change in [C/Fe]
at fixed [Fe/H] results in a slightly cooler main sequence turn-off
point and giant branch but a slightly warmer lower main sequence
\citep{Dotter07,VandenBerg12}.

\section{Data}
\label{s:data}

\subsection{SDSS Data}

The main analysis presented here results from fitting our updated CvD
stellar population model to very high-S/N stacks of passive galaxies,
binned by stellar velocity dispersion.  These galaxies are selected
from the SDSS \citep{York00} Main Galaxy Survey \citep{strauss02} Data
Release 7 \citep{abazajian09}, within a narrow redshift interval
($0.025 < z < 0.06$).  Following the methodology of \citet{peek10}, we
select passive galaxies by requiring that they have no detected
emission in H$\alpha$ nor in [\ionn{O}{ii}]$\lambda$3727.

The individual spectra have moderate S/N, typically $\sim 20$
{\AA}$^{-1}$, so we stack the spectra of hundreds of galaxies to
achieve the very high S/N needed for detailed abundance analysis.
Passive galaxy star formation histories vary strongly as a function of
their stellar velocity dispersion, but also at fixed $\sigma$
depending on how far they scatter off the Fundamental Plane
\citep[FP;][]{graves10, springob12}.  We therefore have chosen to
stack galaxies in bins of $\sigma$, in order to study the main
dimension of stellar population variation, but restrict our sample to
galaxies that lie on the FP (i.e., in the central FP slice defined in
\citealt{graves10}).  The two-dimensional space of galaxy star
formation histories will be explored in a future paper.

We stack the spectra in seven bins of velocity dispersion, with mean
values of $\sigma$ = 88, 112, 138, 167, 203, 246, and 300 km s$^{-1}$.
The effective resolution of the data therefore varies from $R \approx
400-1200$ owing to intrinsic Doppler broadening.  The SDSS spectra are
obtained through 3$''$ diameter fibers, which sample the inner
$\approx0.8\,R_e$ for the smallest $\sigma$ bin and the inner
$\approx0.4\,R_e$ for the largest $\sigma$ bin.  Each individual
spectrum was continuum-normalized and convolved to an effective
dispersion of 350 km s$^{-1}$ before stacking, and each spectrum
contributed equally to the stack.  Problematic pixels (e.g., those
under bright sky lines) were masked before stacking.  The resulting
stacked spectra have S/N at 5000\,\AA\, that ranges from $\approx 500$
{\AA}$^{-1}$ to $\approx 1800$ {\AA}$^{-1}$.  Four wavelength
intervals were defined for the SDSS stacks, 0.40$\mu$m--0.48$\mu$m,
0.48$\mu$m--0.58$\mu$m, 0.58$\mu$m--0.64$\mu$m, and
0.80$\mu$m--0.88$\mu$m for fitting models to data.

\begin{figure}[!t]
\center
\resizebox{3.6in}{!}{\includegraphics{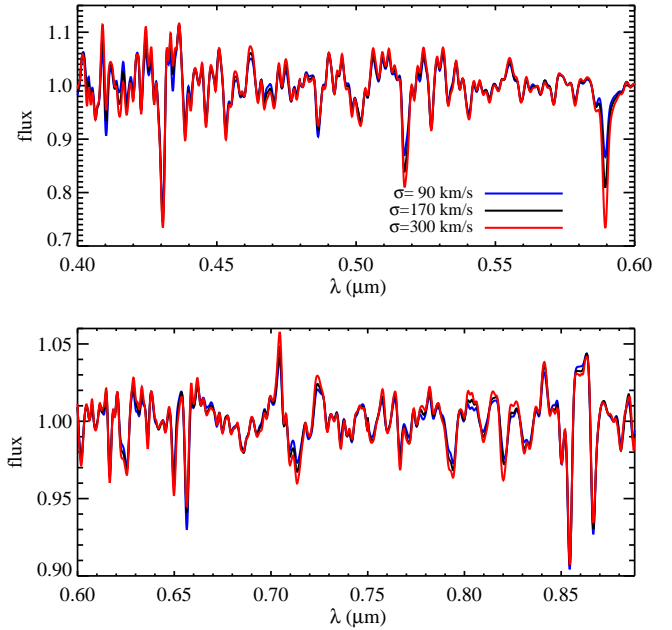}}
\caption{Continuum-normalized stacked spectra of SDSS early-type
  galaxies in three velocity dispersion bins.}
\label{fig:data}
\end{figure}

In Figure \ref{fig:data} we show the stacked spectra in three $\sigma$
bins.  The fluxes have been continuum normalized in order to focus
attention on the differences in the line-strengths between different
stacks.  Notice that the differences are relatively subtle, with
features strengths varying by only a few percent from the lowest to
the highest dispersion bins.

\begin{figure*}
\center
\resizebox{7in}{!}{\includegraphics{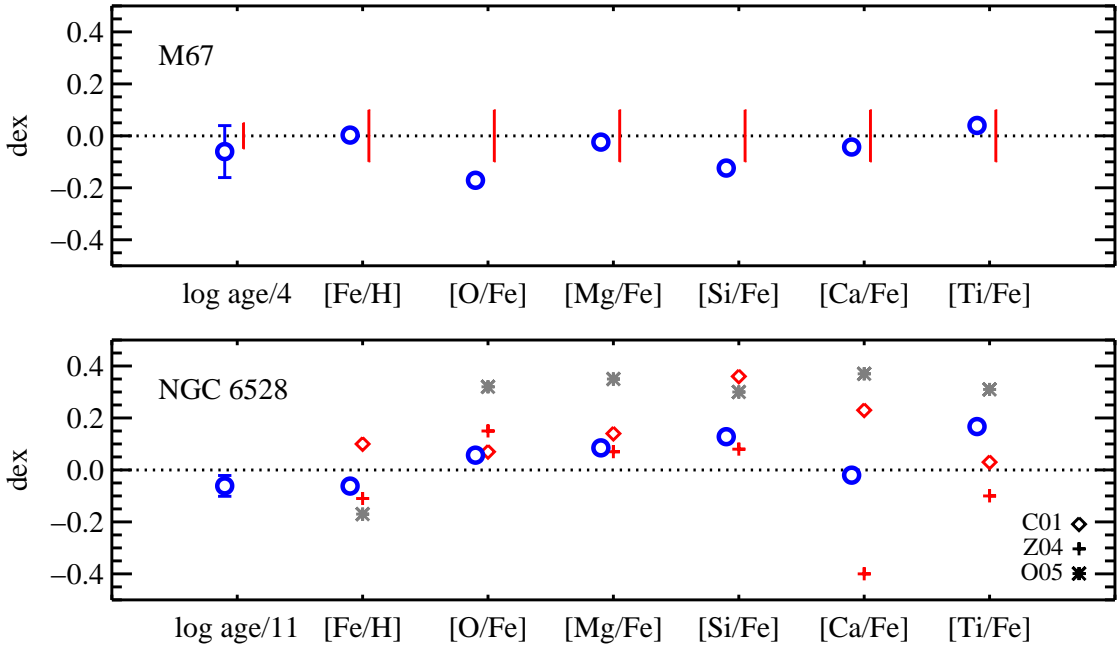}}
\caption{Best-fit parameters for the star clusters M67 (top panel) and
  NGC 6528 (bottom panel).  Results from our full-spectrum fitting of
  the integrated light (open blue circles) are compared to literature
  values based on high resolution spectra of individual member stars
  (red lines and symbols).  Fiducial ages for M67 and NGC6528 are
  taken to be 4 and 11 Gyr, respectively, based on isochrone fitting
  to their CMDs.  Abundance patterns for NGC 6528 are from
  \citet[][C01]{Carretta01}, \citet[][Z04]{Zoccali04}, and
  \citet[][O05]{Origlia05}. For M67 we adopt a fiducial abundance
  pattern of $0.0\pm0.1$ \citep[roughly $1\sigma$
  errors][]{Tautvaisiene00, Shetrone00}.  Overall the agreement
  between our model predictions and the literature values is good,
  although the large variation in the literature determination of many
  of the elemental abundances for NGC 6528 prohibits a more definitive
  test of our model.}
\label{fig:globtest}
\end{figure*}

\begin{deluxetable*}{ccccccccccccccccccc}
\tablecaption{Results from Modeling Stacked SDSS Early-Type Galaxy Spectra}
\tablehead{ \colhead{log $\sigma$} & \colhead{log $M_\ast$} & \colhead{rms} & \colhead{age} &
  \colhead{log $f_y$} & \colhead{[Fe/H]} & \colhead{C} & \colhead{N}&\colhead{O}&
\colhead{Na} & \colhead{Mg} & \colhead{Si} &
\colhead{Ca}  & \colhead{Ti}  & \colhead{V} &
\colhead{Cr} & \colhead{Mn}  & \colhead{Co}  &
\colhead{Ni} \\ 
\colhead{$\kms$} & \colhead{$\Msun$} & \colhead{\%} & \colhead{Gyr} & 
\colhead{} & \colhead{}&\colhead{}&
\colhead{} & \colhead{} & \colhead{} & \colhead{} & \colhead{} &
\colhead{}  & \colhead{}  & \colhead{} &
\colhead{} & \colhead{}  & \colhead{}  & \colhead{} }
\startdata
 1.94 & 9.63 & 0.22 &  6.2 &  -0.7 & -0.07 &  0.05 &  0.02 &  0.03 & -0.16 &  0.05 &  0.02 &  0.03 &  0.05 &  0.03 &  0.02 &  0.02 &  0.03 & -0.00  \\
 2.05 & 9.75 & 0.22 &  6.5 &  -0.9 & -0.05 &  0.07 &  0.08 &  0.06 & -0.09 &  0.08 &  0.00 &  0.03 &  0.07 &  0.01 &  0.00 &  0.03 &  0.06 & -0.00 \\
 2.14 & 9.80 & 0.21 &  6.8 &  -2.2 & -0.04 &  0.09 &  0.13 &  0.10 &  0.02 &  0.09 &  0.03 &  0.03 &  0.08 &  0.03 & -0.01 &  0.03 &  0.10 &  0.01  \\
 2.23 &10.08 & 0.22 &  7.0 &  -4.1 & -0.03 &  0.13 &  0.18 &  0.17 &  0.12 &  0.12 &  0.05 &  0.03 &  0.09 &  0.03 & -0.01 &  0.03 &  0.16 &  0.02  \\
 2.30 &10.55 & 0.26 &  7.6 &  -4.3 & -0.01 &  0.16 &  0.21 &  0.20 &  0.22 &  0.15 &  0.09 &  0.03 &  0.11 &  0.03 & -0.03 &  0.02 &  0.20 &  0.03  \\
 2.39 &10.70 & 0.29 & 11.0 &  -4.3 & -0.02 &  0.19 &  0.26 &  0.25 &  0.33 &  0.20 &  0.13 &  0.04 &  0.12 & -0.01 & -0.02 &  0.05 &  0.27 &  0.01  \\
 2.47 &11.07 & 0.34 & 11.8 &  -4.4 &  0.00 &  0.21 &  0.27 &  0.28 &  0.43 &  0.22 &  0.16 &  0.02 &  0.12 & -0.02 & -0.03 &  0.05 &  0.26 &  0.07 \\

\enddata
\vspace{0.1cm} 
\tablecomments{Stellar masses are averages within the $\sigma$ bins
  and they assume a \citet{Kroupa01} IMF.  The rms deviation between
  the model and data is computed over the full wavelength range used
  in the fit.  Ages are light-weighted and the fraction of mass in the
  3 Gyr component, $f_y$, is included in the table.  Abundances of
  elements are quoted as [X/Fe].  Formal statistical errors are less
  than 1\% and so are omitted from the table.  Systematic errors
  dominate the error budget, and are estimated to be $0.05$ dex or
  less.  See Section \ref{s:sys} for further discussion.}
\label{t:res}
\end{deluxetable*}

\subsection{Star Cluster Data}

In this paper we will also analyze integrated light spectra of
metal-rich star clusters in order to test the model.  We focus on two
$\sim Z_\Sol$ clusters, M67 and NGC 6528.  The former is an
intermediate-age open cluster with solar-scaled abundance ratios,
while the latter is an old bulge globular cluster with enhanced
abundance ratios.  For M67 we adopt an age of 4 Gyr derived from its
color-magnitude diagram \citep[CMD;][]{vandenberg07, Magic10},
[Fe/H]=0.0, and solar-scaled abundance ratios with an uncertainty of
0.1 dex \citep{Tautvaisiene00, Shetrone00}.  For NGC 6528 we adopt an
age of 11 Gyr, consistent with recent CMD-based determinations
\citep{Feltzing02, Momany03}.  The elemental abundance data are
derived from high-resolution spectra of individual RGB stars, as
reported in \citet{Carretta01}, \citet{Zoccali04}, and
\citet{Origlia05}.

The integrated light spectrum for M67 was constructed in
\citet{Schiavon04} by combining spectra of individual member stars
with a \citet{Salpeter55} IMF.  M67 contains a large population of
blue straggler stars, but these were omitted in the integrated
spectrum.  The spectrum covers the range 3640\AA, to 5400\AA, at a
resolution of 2.7\AA.  The S/N of the integrated spectrum was not
provide by Schiavon et al., so we adopt a nominal S/N of 100 per
pixel.

For NGC 6528 we use the integrated light spectrum from the library of
globular cluster spectra presented in \citet{Schiavon05}.  These
spectra were obtained via drift scan observations, so the spectra
reflect the true integrated light of the clusters.  The spectrum
covers the wavelength range 3360\AA\, to 6430\AA\, at a resolution of
$\sim3.1$\AA.  The S/N ranges from $\approx$50 per pixel at 4000\AA\,
to 150 per pixel at 5000\AA.  In this paper we analyze the spectrum
extracted within the central $7\arcsec$.  Several spectral regions had
to be masked owing to sub-optimal sky subtraction and bad CCD columns.
These regions include $4155$\AA-$4165$\AA, $4535$\AA-$4565$\AA,
$4850$\AA-$4870$\AA, and $5035$\AA-$5065$\AA.

For both spectra we use two wavelength intervals when fitting the
model to these data: $4000$\AA-$4600$\AA, and $4600$\AA-$5350$\AA.

\begin{figure*}[!t]
\center
\resizebox{7in}{!}{\includegraphics{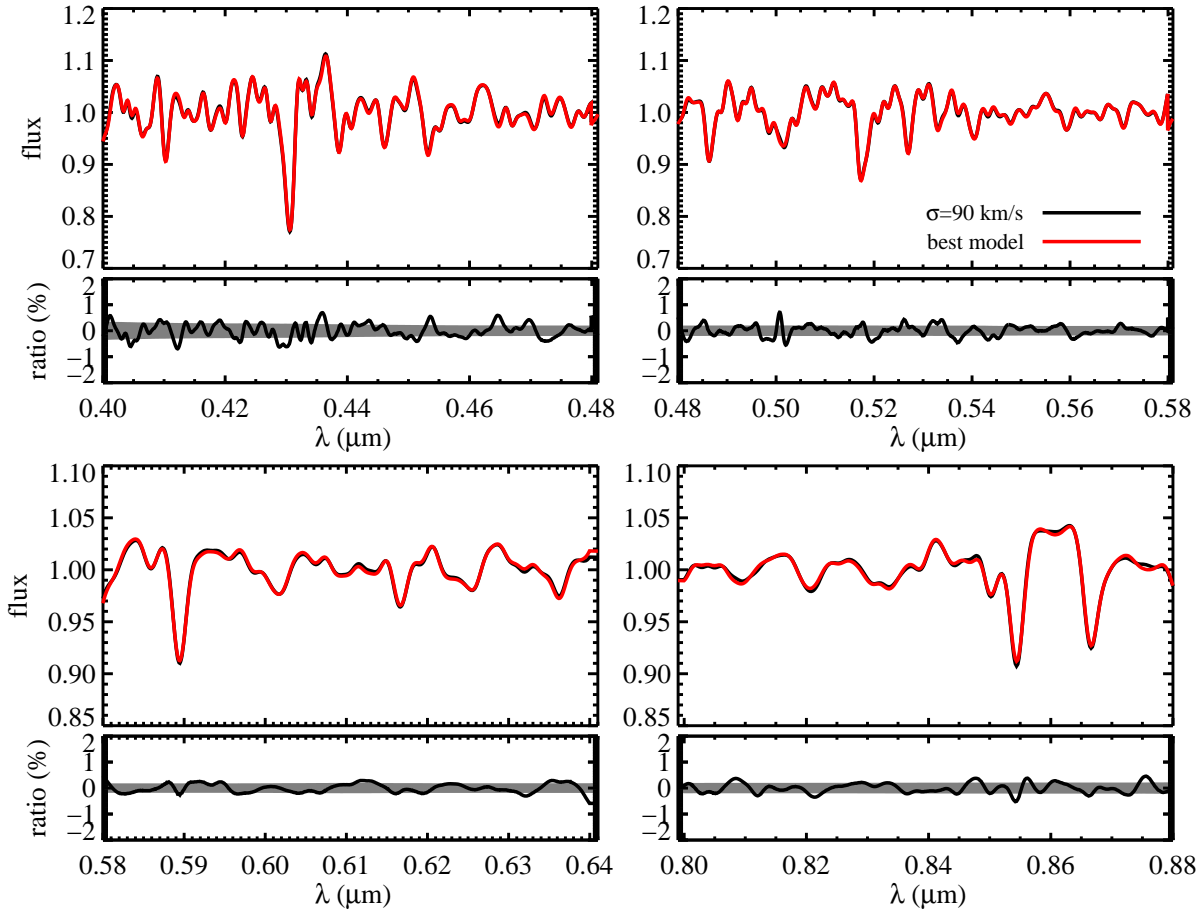}}
\caption{Comparison between the SDSS early-type galaxy stacked
  spectrum in the lowest velocity dispersion bin (black lines) and the
  best-fit model (red lines).  Within each wavelength range, the top
  panel shows the continuum-normalized fluxes and the lower panel
  shows the ratio between data and model.  The grey band denotes the
  $1\sigma$ error limits of the data.  Overall the quality of the fit
  is excellent.}
\label{fig:res1}
\end{figure*}

\begin{figure*}[!t]
\center
\resizebox{7in}{!}{\includegraphics{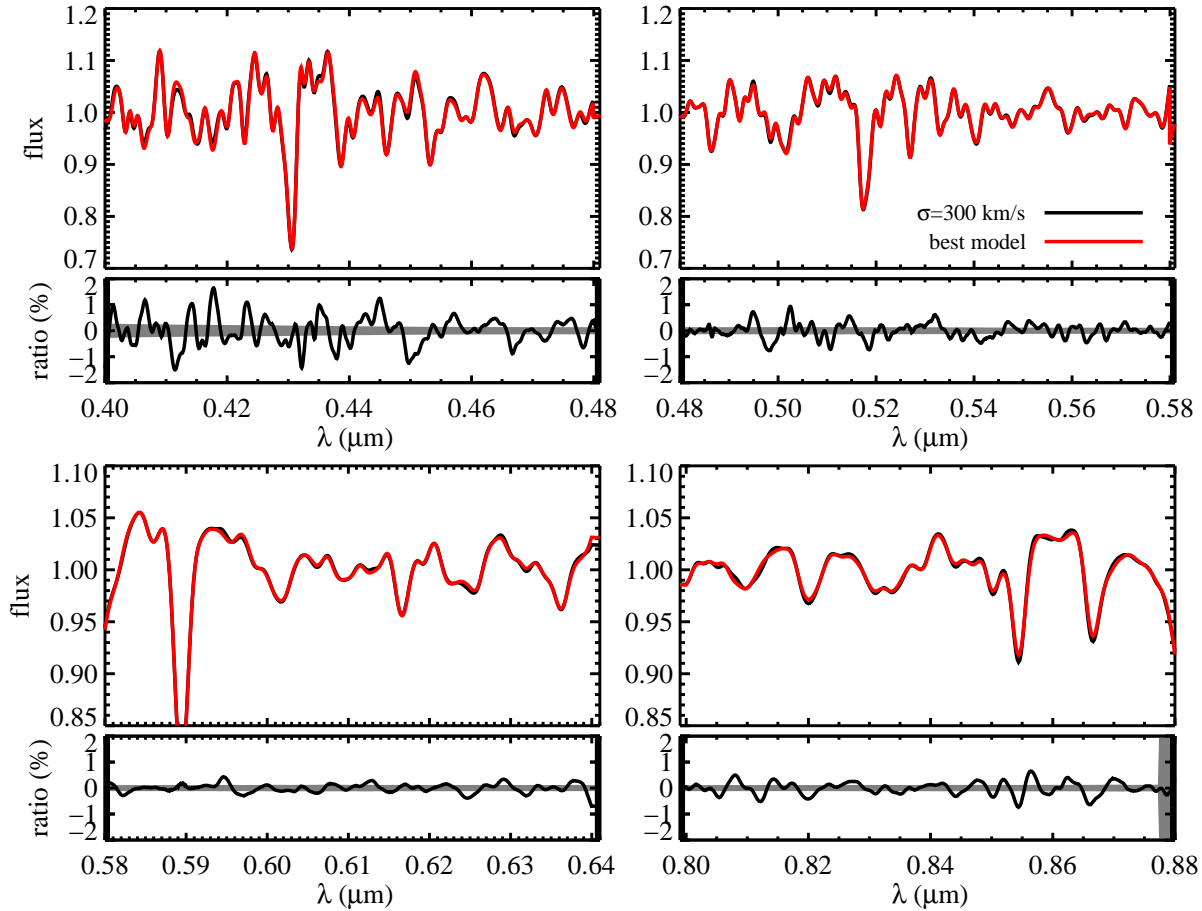}}
\caption{Same as Figure \ref{fig:res1}, now showing results for the
  highest velocity dispersion bin.  As in Figure \ref{fig:res1} the
  overall quality of the fit is excellent.}
\label{fig:res2}
\end{figure*}

\begin{figure*}[!t]
\center
\resizebox{6.5in}{!}{\includegraphics{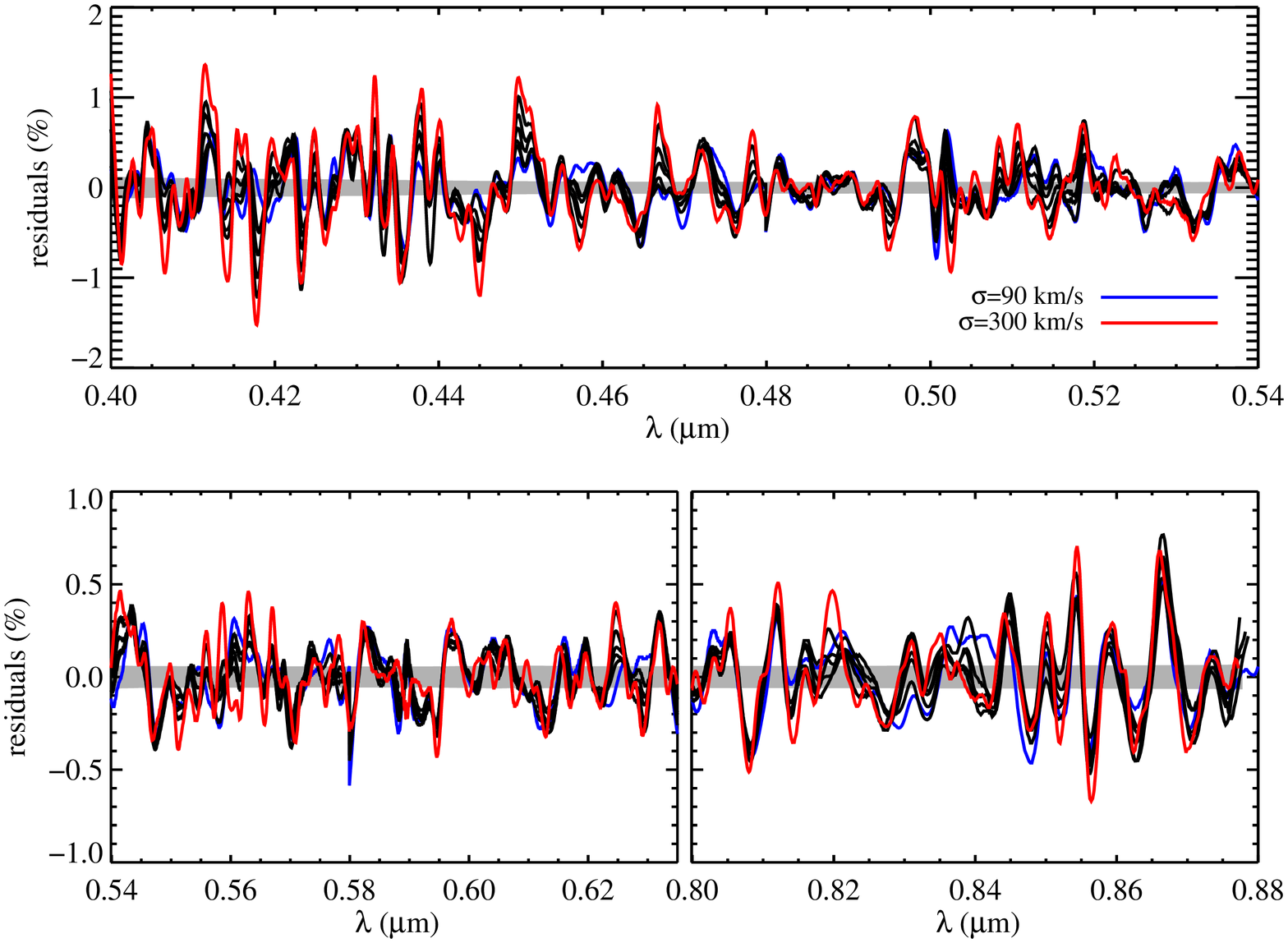}}
\caption{Residuals between the data and best-fit models for all seven
  stacked early-type galaxy spectra analyzed in this paper.  The
  shaded region marks the noise limit of the middle $\sigma$ bin.  In
  most spectral regions the residuals are constant across the sample,
  which is an important result because it means that the higher
  $\sigma$ galaxies, which harbor abundance patterns significantly
  different from the solar neighborhood, are nearly as well described
  by our model as the lower $\sigma$, less metal-enhanced galaxies.
  In a few spectral regions, e.g., at $0.45\mu m$ the residuals
  correlate with galaxy velocity dispersion, indicating that our model
  is not capturing the totality of variation in the observed spectra.}
\label{fig:resall}
\end{figure*}

\section{Testing The Model Against Metal-Rich Star Clusters}
\label{s:calib}

Before discussing the main results of this paper, in this section we
present the results of fitting our models to spectra of the metal-rich
star clusters M67 and NGC 6528.  Star clusters are of course simpler
to model than galaxies, owing to the fact that they are at least
approximately coeval and monometallic.  They therefore can serve as
useful tests of any SPS model.

In Figure \ref{fig:globtest} we compare the best-fit ages, [Fe/H], and
several $\alpha$ elements for these two clusters.  Turning first to
M67, we derive an age of $3.5\pm0.3$ Gyr, in excellent agreement with
the CMD-based age.  We also find excellent agreement for Fe, Mg, Ca,
and Ti.  The abundances of O and Si are underestimated by $1-2\sigma$.
Because the effects of O and Si on the spectrum are somewhat more
subtle than the other $\alpha$ elements, integrated light spectra that
include the redder spectral coverage for M67 are needed before a more
definitive conclusion regarding these elements can be reached.  Though
not shown in Figure \ref{fig:globtest}, we also find [C/Fe]$=-0.06$
and [N/Fe]$=-0.02$ for M67, which agrees within 0.1 dex with
literature determinations of C and N from main sequence turn-off stars
in M67 \citep{Shetrone00}.

The model also provides an excellent fit to the age and [Fe/H]
abundance of NGC 6528.  The $\alpha$ elements also broadly agree with
the literature data, although perhaps the most striking impression
from Figure \ref{fig:globtest} is the large scatter between the
literature estimates for the abundance ratios.  The most dramatic
difference is in the estimated [Ca/Fe] abundance, where
\citet{Origlia05} find [Ca/Fe]$=+0.4$ while \citet{Zoccali04} quote
[Ca/Fe]$=-0.40$; a difference of 0.8 dex!  We have plotted the
\citet{Origlia05} results in grey because these authors model
high-resolution NIR spectra while \citet{Carretta01} and
\citet{Zoccali04} model high-resolution blue spectra.  We also draw
attention to the fact that these two authors find relatively modest
enhancements in O and Mg, less than 0.2 dex.  Thus, while NGC 6528 is
$\alpha-$enhanced, it is not tremendously so.

For NGC 6528 we do not compare results for C nor N because these
elements can be synthesized in the envelopes of red giant branch (RGB)
stars, so abundances of these elements derived from such stars do not
reflect the original (i.e., main sequence) abundances.  Our model
recovers the latter, whereas literature estimates of C and N for this
cluster are based on RGB stars.  Owing to the large discrepancy in
reported literature abundances, even for important species such as Fe,
O, Mg, Si, Ca, and Ti, we made no effort to compare to other elements
as their derived abundances from high-resolution stellar spectroscopy
appear to be even more uncertain.

Our model successfully recovers the ages, [Fe/H], and $\alpha$ element
abundances of metal-rich star clusters.  Further progress on testing
SPS models must await 1) higher quality integrated light spectra of
star clusters covering a wider wavelength range and at higher S/N than
is presently available; 2) a thorough evaluation of the results from
high-resolution stellar spectroscopy.  Until a consensus emerges on
the abundance patterns of the stars in metal-rich star clusters, such
systems cannot be used to rigorously test abundance ratio predictions
of SPS models.

\section{Results}
\label{s:res}

We now turn to our main results --- the analysis of high quality
stacked spectra of early-type galaxies from the SDSS.  In this section
we present and discuss the overall quality of the fits, the detailed
abundance patterns, and the derived effective temperatures of the
stars.  Table \ref{t:res} presents the main derived parameters from
modeling the stacked spectra.  Results for the abundance ratios of
neutron capture elements Sr and Ba were presented in
\citet{Conroy13a}, and the results for the variation in the IMF with
galaxy $\sigma$ will be presented in future work.  We simply note here
that the IMF trends with $\sigma$ are similar to those presented in
\citet{Conroy12b}.  We include in this table estimated stellar masses
based on the mean $r-$band luminosities of the galaxies in each bin
combined with the derived stellar $M/L$ values assuming a
\citet{Kroupa01} IMF.

\subsection{Overall Quality of the Fits \& Uniqueness of the Derived
  Parameters}

We begin by discussing the overall quality of the model fits.  In
Figures \ref{fig:res1} and \ref{fig:res2} we show the data and
best-fit model for the lowest and highest $\sigma$ bins.  Within each
panel the fluxes have been continuum-normalized by a high-order
polynomial.  These figures demonstrate the overall very high quality
of the fits.  In the lowest dispersion bin the rms between data and
model is only 0.22\%, and $\chi^2$/dof$=1.1$, indicating that our
model is able to fully capture the observed variation in the spectrum
to the precision allowed by the data.  At the highest dispersion bin,
where the abundance ratios deviate most strongly from the solar-scaled
values, the rms is 0.34\%, with $\chi^2$/dof$=4.7$.  It is not
surprising that the fit is formally worse for the highest dispersion
bin, both because the S/N is higher, which imposes more stringent
demands on the model, and because the strong deviations from
solar-scaled abundance ratios means that the synthetic stellar spectra
play an important role in the modeling.  The synthetic spectra carry
significant uncertainties (e.g., due to the treatment of the model
atmospheres and incomplete line lists), so their greater weight in the
model will tend to result in larger model uncertainties.

In Figure \ref{fig:resall} we show the residuals now for all seven
bins.  The most striking feature of the residuals is their similarity
across the sample.  For example, the residuals around the
\ionn{Ca}{ii} triplet at $8500$\AA-$8700$\AA\, are relatively
significant, but the magnitude of the residuals does not vary strongly
with galaxy velocity dispersion \citep[see also][where this was
demonstrated on an object-by-object basis]{Conroy12b}.  This weak
dispersion-dependence in the residuals is important for several
reasons. First and foremost, it suggests that the trends in the
derived parameters, to be discussed later in this section, are likely
insensitive to remaining model systematics.  In addition, the fact
that the residuals at high dispersion are similar to the lowest
dispersion bin, where the abundance ratios are close to solar-scaled
values, suggests that the model deficiencies lie in the base models
constructed with empirical stellar spectra.  It is worth pointing out
that the MILES stellar library, which forms the core of the model in
the optical, has typical S/N of $\sim150$ \AA$^{-1}$, so at the level
of interest here the uncertainties in the empirical spectra probably
play an important role (the residuals in Figure \ref{fig:resall} are
smooth in part because the model and data have been smoothed to
$\sigma=350\kms$).

Some systematics with dispersion are evident, for example at 4500\AA\,
and 4675\AA.  These are probably due to features that are not
well-captured in our synthetic spectral library.  In future work we
will investigate the sources of line opacity in these wavelength
regions in an attempt to further improve the models.

It is also apparent from Figure \ref{fig:resall} that the residuals
are larger in the blue than in the red.  This is likely due to the
fact that the blue spectral region contains more and stronger atomic
and molecular absorption lines than the red (e.g., compare Figures
\ref{fig:blue} and \ref{fig:red}).  Galaxies with non-solar abundance
patterns will therefore place more demanding requirements on our
models in the blue compared to the red.  It is also evident that the
residuals in the blue are much more dependent on the galaxy velocity
dispersion than in the red.  In the red spectral region the median
absolute residuals are approximately constant with $\sigma$ at 0.1\%
while in the blue they increase from 0.16\% at low $\sigma$ to 0.25\%
at high $\sigma$ (cf. Figures \ref{fig:res1} and \ref{fig:res2}).

Our model contains 40 free parameters that are fit to the data via an
MCMC algorithm.  The parameter space is very large and it is difficult
with any algorithm to ensure that one has reached the true global
maximum of the likelihood surface.  Nonetheless, we have performed a
large number of tests, including re-starting the chains at random
locations in parameter space.  Each time the chain re-converges to the
same maximum.  Moreover, tests have demonstrated that the likelihood
surface is very smooth, especially for the high S/N data we are
modeling, and so for this reason isolated local maxima appear to be
rare.

\begin{figure}
\center
\resizebox{4in}{!}{\includegraphics{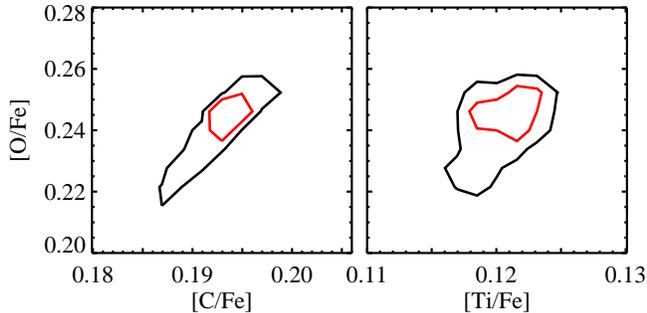}}
\caption{Covariance between [O/Fe], [C/Fe] and [Ti/Fe] for the SDSS
  stacked spectrum with log $\sigma=2.39\kms$.  Contours show 68\% and
  95\% confidence limits in the posterior distribution.  The response
  function due to O variation is similar (though opposite in sign) to
  C variation at $\lambda<6000$\AA\, (see Figure \ref{fig:blue}),
  which is mostly a consequence of molecular equilibrium involving CO.
  Similarly, in the red, the response functions of O and Ti resemble
  one another, here because of TiO.  This figure demonstrates that, in
  the context of our model, we are able to separately constrain C, O,
  and Ti very well (notice the small range of the $x$ and $y$-axes).}
\label{fig:degen}
\end{figure}

We can also explore the level of covariance (degeneracy) between
various parameters.  This was discussed in \cite{Conroy12b}; we
revisit this issue here in a different context.  In Figure
\ref{fig:degen} we plot the covariance between the abundances of C, O,
and Ti, as the abundances of three elements may on physical grounds be
expected to display some degree of covariance.  For example,
inspection of Figures \ref{fig:blue} and \ref{fig:red} demonstrates
that the response functions of these elements are similar in several
respects.  This similarity is driven in part by molecular equilibrium
involving CO and TiO.  But from Figure \ref{fig:degen} it is evident
that our model is capable of separately measuring these three
quantities with very high accuracy and modest levels of degeneracy
(basically no degeneracy at 68\% confidence and a moderate degree of
degeneracy at the 95\% confidence limit for [O/Fe] vs. [C/Fe], as
might be expected from Figure \ref{fig:blue}).  We have inspected a
wide variety of other correlations and find in all cases the level of
degeneracy between parameters to be very small.  When fitting spectra
with very high S/N there is apparently sufficient information to
strongly constrain at least 40 free parameters.  In this regime it is
the systematic uncertainties that dominate the error budget, the
magnitude of which we explore in Section \ref{s:sys}.

\begin{figure*}[!t]
\center
\resizebox{3.5in}{!}{\includegraphics{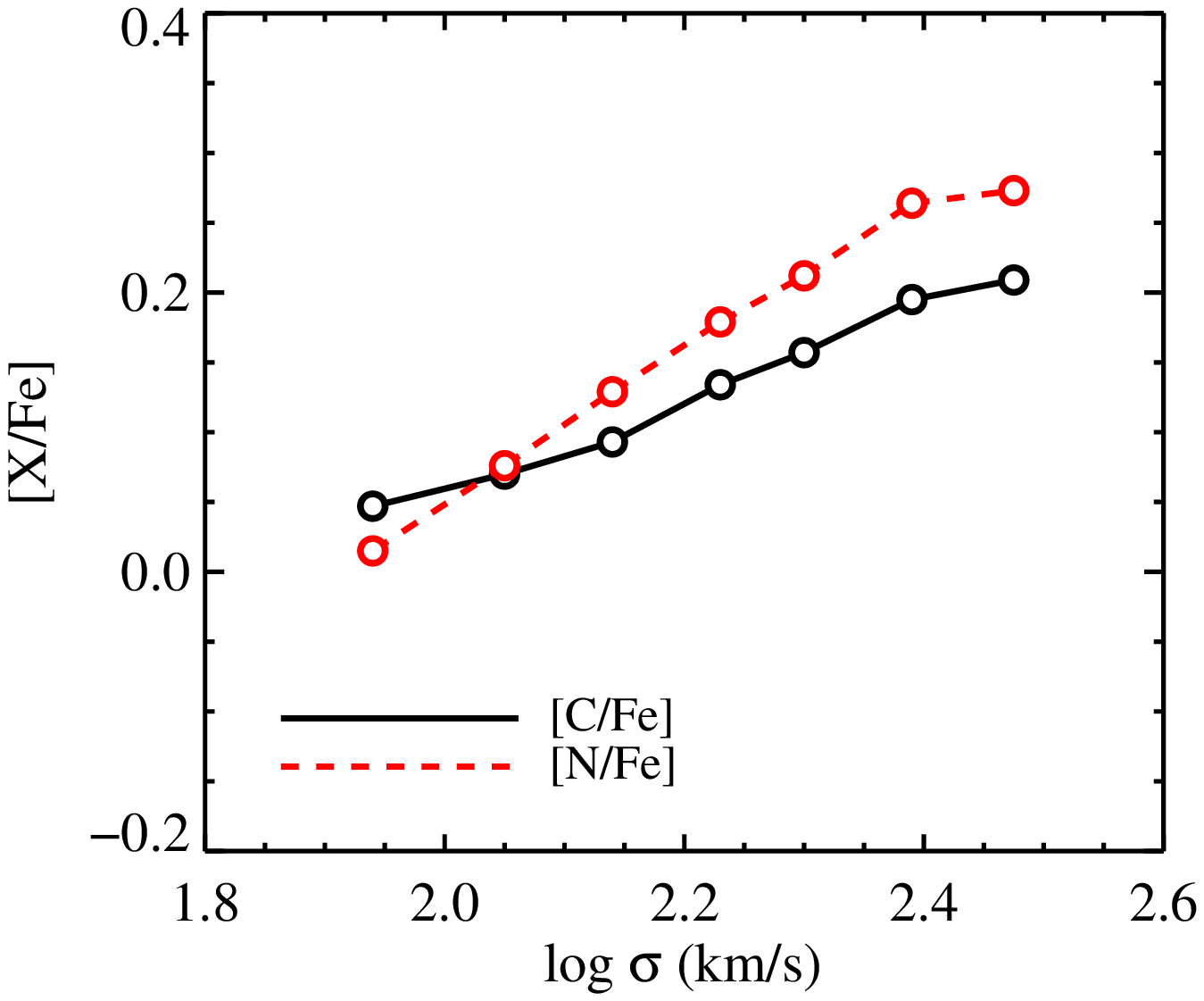}}
\resizebox{3.5in}{!}{\includegraphics{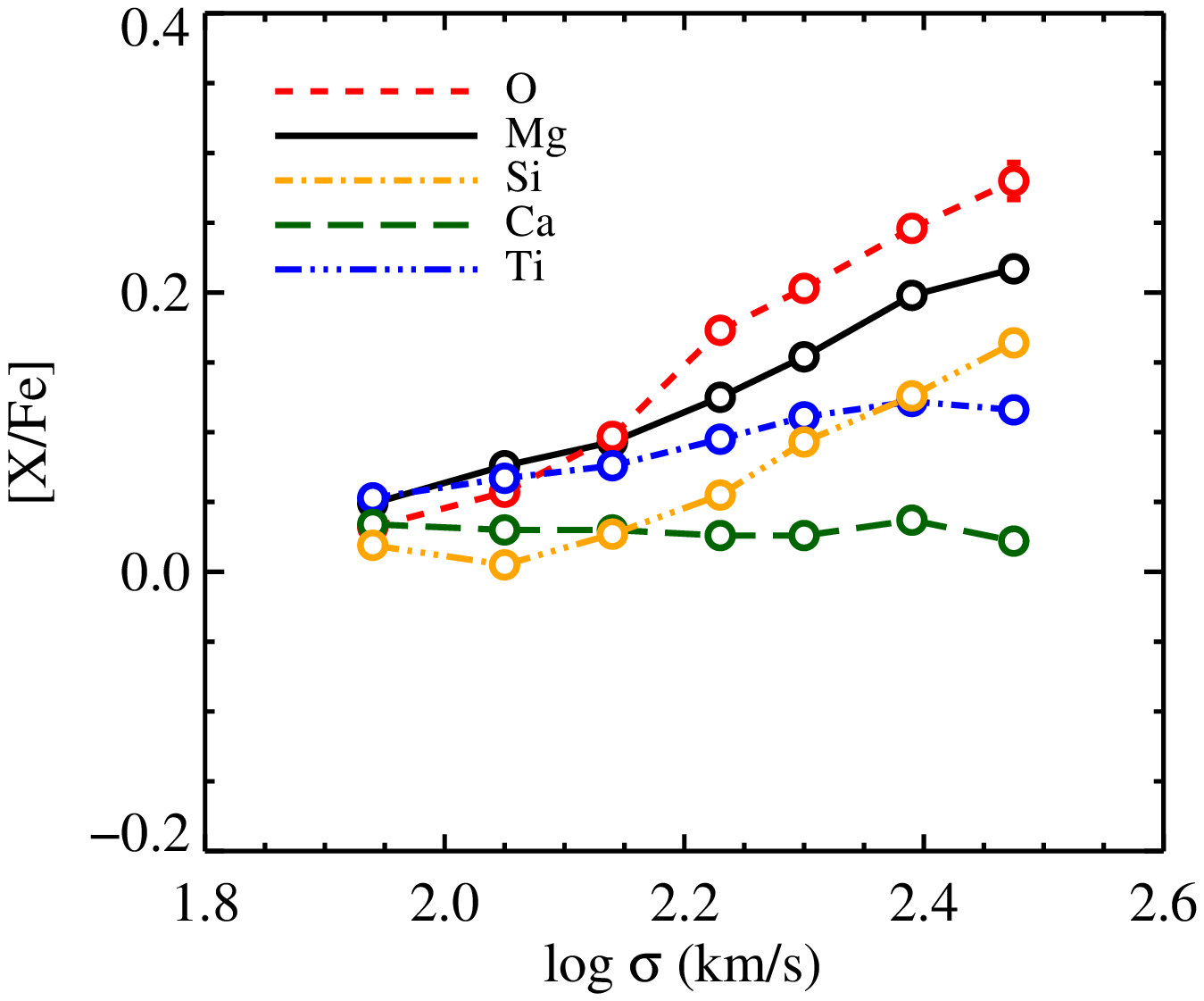}}
\caption{Abundance ratios of C, N, and the $\alpha$ elements as a
  function of early-type galaxy velocity dispersion.  In most cases
  the statistical errors are smaller than the symbol sizes.  C, N, O,
  and Mg all seem to track each other fairly closely, while the
  heavier $\alpha$ elements Si, Ca, and Ti show weaker trends with
  $\sigma$.  Ca clearly tracks Fe instead of O.  Systematic errors are
  probably $<0.05$ dex (see Section \ref{s:sys}).}
\label{fig:cn_alpha}
\end{figure*}

\begin{figure*}[!t]
\center
\resizebox{3.5in}{!}{\includegraphics{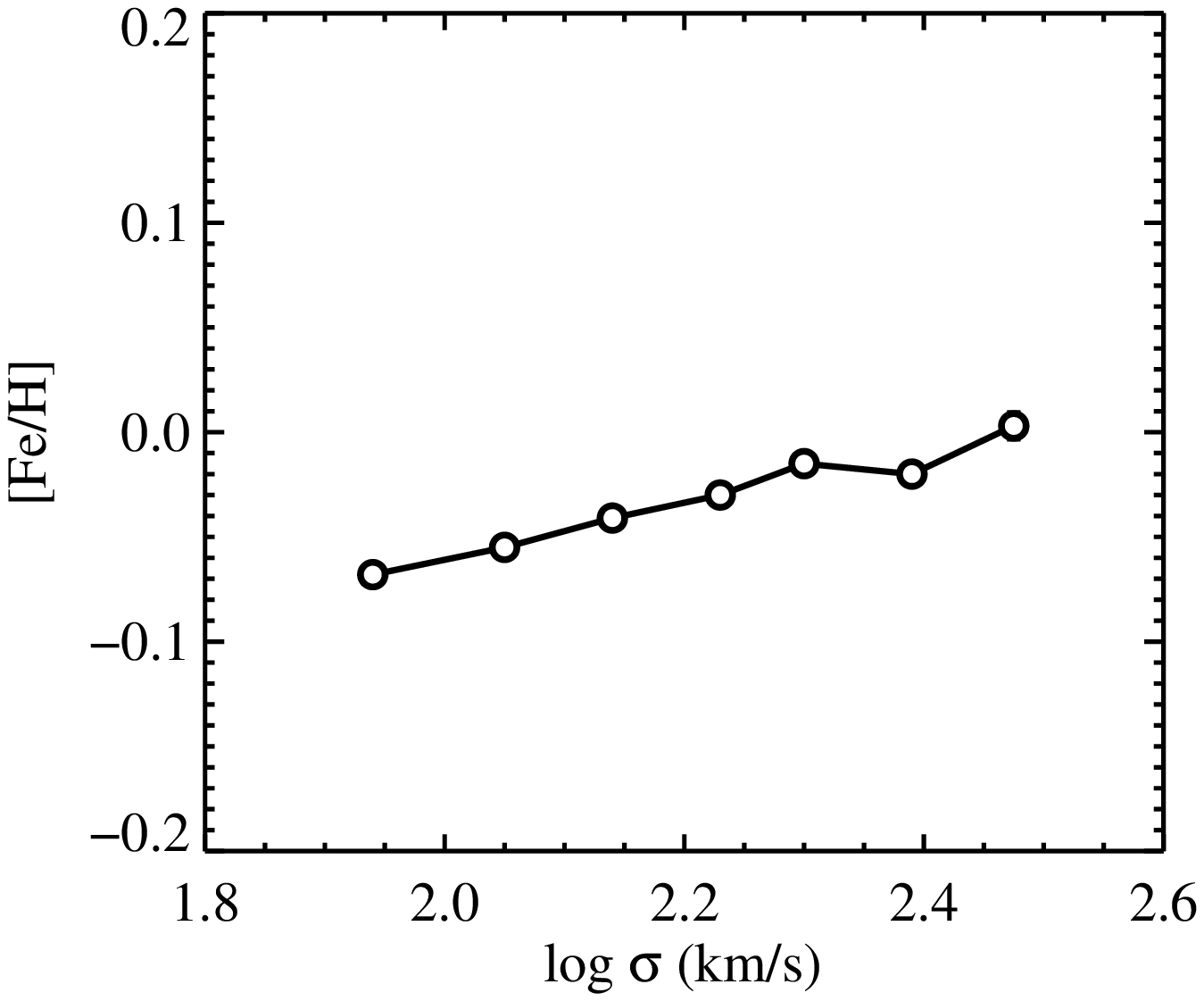}}
\resizebox{3.5in}{!}{\includegraphics{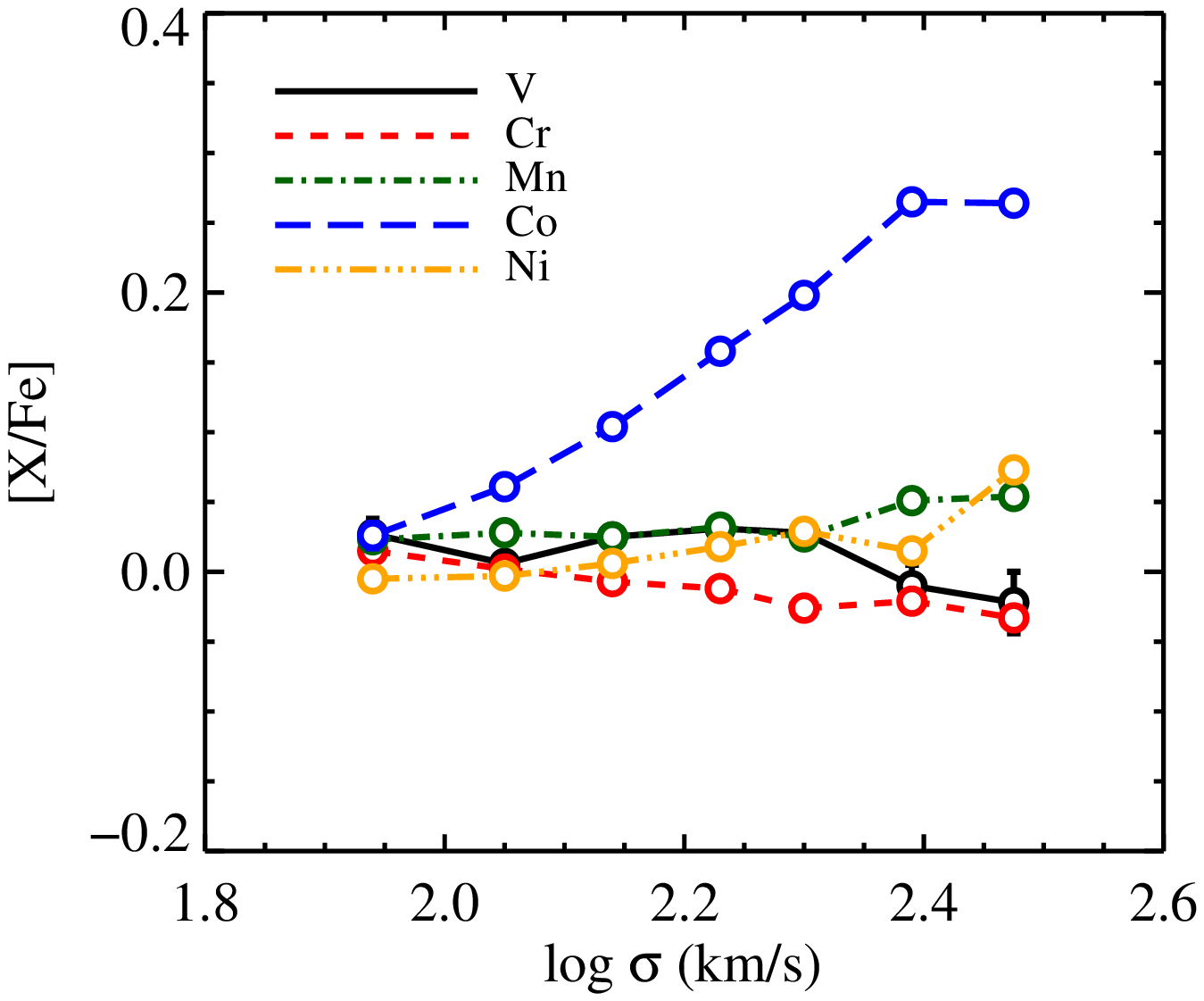}}
\caption{Abundance ratios of the iron peak elements as a function of
  early-type galaxy velocity dispersion.  In most cases the
  statistical errors are smaller than the symbol sizes.  All iron peak
  elements track Fe except for Co, which seems to track the light
  $\alpha$ elements O and Mg.  Systematic errors are probably $<0.05$
  dex (see Section \ref{s:sys}).}
\label{fig:fe_fepeak}
\end{figure*}

\subsection{Abundance Patterns}

We now turn to the elemental abundance ratios derived from our sample
of stacked SDSS early-type galaxy spectra.  

In Figure \ref{fig:cn_alpha} we show the relations between velocity
dispersion and C and N (left panel), and the $\alpha$ elements O, Mg,
Si, Ca, and Ti (right panel).  In most cases the formal statistical
errors are smaller than the symbol sizes.  The abundance of both C and
N increase from [C/Fe]$\approx$[N/Fe]$\approx0.0$ at low $\sigma$ to
$0.2-0.3$ at high $\sigma$.  This is in qualitative agreement with
previous results \citep{Kelson06, Graves07, Schiavon07, Smith09,
  Johansson12}.

The derived abundance ratio patterns of the $\alpha$ elements show
several interesting features.  First, all ratios are close to the
solar-scaled values at low $\sigma$.  Second, the trends with $\sigma$
are stronger for the lighter elements O and Mg compared to the heavier
elements Si, Ca, and Ti.  Mg appears to track O closely, with a slight
preference for [Mg/O]$<0.0$ at high $\sigma$.  Ca tracks Fe closely
over the full sample, in agreement with previous work \citep{Saglia02,
  Cenarro03, Thomas03b, Graves07, Schiavon07, Smith09, Johansson12}.
This suggests that a substantial amount of Ca is formed in Type Ia
SNe, as expected from theoretical yields \citep{Nomoto84}.  The
enhancement of Si and Ti at high $\sigma$ also broadly agrees with
earlier work \citep{Milone00, Johansson12, Worthey13}.

In Figure \ref{fig:fe_fepeak} we show the results for [Fe/H] in
addition to the other iron peak elements V, Cr, Mn, Co, and Ni, all of
which are measured here for the first time.  The [Fe/H] abundance
increases modestly over the sample by only 0.07 dex.  This remarkable
uniformity of the Fe abundance, in light of the substantial variation
in the abundance of the light and $\alpha$ elements, poses a challenge
to chemical evolution models of early-type galaxies.  The abundances
of the iron peak elements V, Cr, Mn, and Ni closely track Fe over the
full sample, implying that these elements form in the same
nucleosynthetic sites.  In stark contrast is Co, which increases in
relative abundance to Fe as $\sigma$ increases, reaching
[Co/Fe]$\approx0.27$ at the highest dispersions.

This result is explored further in Figure \ref{fig:cnco}, where we
show the abundance ratios of several elements with respect to O as a
function of $\sigma$.  Clearly both N and Co track O remarkably
closely.  C also tracks O more closely than Fe, but [C/O] is not quite
as constant as [N/O] and [Co/O], suggesting that C and N may not be
forming in exactly the same environments.

Although not shown in a figure, the [Na/Fe] abundance also varies
considerably across the sample, from $-0.16$ to $0.43$, qualitatively
tracking the light $\alpha$ elements (see Table \ref{t:res}).  In
detail however the Na abundance variation is considerably stronger
than the other elements.  The only two Na lines available to us are
the \ionn{Na}{i} features at 5895\AA\, and 8190\AA, respectively.
The former is well-known to be affected by the presence of an
interstellar medium (ISM).  However, the sample was selected to have
no detectable emission in H$\alpha$ nor in [\ionn{O}{ii}], suggesting
that contamination by ISM absorption should be small.  Moreover, one
might have expected ISM contamination to be greater at lower $\sigma$,
because low mass galaxies are younger, and yet the derived [Na/Fe] at
low $\sigma$ is {\it lower} than either [O/Fe] or [Mg/Fe] (the
presence of an ISM would cause stronger absorption at 5895\AA\, and
thus larger derived [Na/Fe] values).  Of course, it is not necessary
that [Na/Fe] track the other elements in detail, as its
nucleosynthetic origins are quite complex, including contributions
from both massive stars and intermediate-mass asymptotic giant branch
(AGB) stars.  Indeed, the chemical evolution model of
\citet{Arrigoni10} finds a variation in [Na/Fe] in broad agreement
with what we find herein, ranging from $-0.2$ at low mass to $+0.3$ at
high mass.

\subsection{Effective Temperatures}

As discussed in Section \ref{s:model}, one of the free parameters in
our model is the shift in $\teff$ from a fiducial solar metallicity
isochrone.  This is an important parameter to include because, as we
saw in the previous section, the metallicities and abundance patterns
vary by factors of several across our sample.  A change in metallicity
and/or abundance pattern will induce a change in $\teff$ of the stars
\citep{Dotter07}.  It would be difficult to include this variation
self-consistently owing to the large grid of isochrones that would be
required.  The parameter $\Delta \teff$ is meant to capture all of
these effects into a single number.

\begin{figure}
\center
\resizebox{3.5in}{!}{\includegraphics{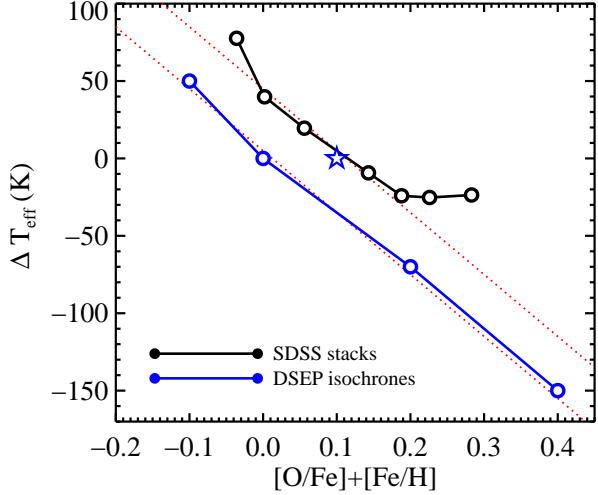}}
\caption{Relation between offset in isochrone $\teff$ and metal
  abundance.  Results are shown for the stacked SDSS early-type galaxy
  data and compared to expectations based on theoretical isochrones
  from the Dartmouth Stellar Evolution Database (DSEP).  For the
  latter, all of the $\alpha$ elements track each other.  The
  theoretical offsets are derived from Figure \ref{fig:iso}. The star
  symbol corresponds to the model with [Fe/H]$=-0.1$ and
  [$\alpha$/Fe]=$+0.1$ and highlights the fact that $\Delta \teff$ is
  not a simple function of [Z/H].  The dotted lines are offset from
  each other by 40K.  The excellent agreement in the {\it slope}
  between our modeling and the DSEP predictions is a strong indication
  that our model is able to reliably measure the distribution of
  $\teff$ in early-type galaxies.}
\label{fig:teff}
\end{figure}

In Figure \ref{fig:teff} we show the resulting relation between
$\Delta \teff$ and the sum of [O/Fe] and [Fe/H], which is a proxy for
[Z/H] \citep[e.g.,][]{Trager00a}.  The relation derived for our sample
of SDSS early-type galaxies is compared to theoretical expectations
from the Dartmouth isochrones \citep{Dotter08b}.  For the latter, all
of the $\alpha$ elements track each other. The early-type galaxy
sample increases in $\sigma$ from low to high metallicity. The
theoretical predictions are based on the models shown in Figure
\ref{fig:iso}.  The star symbol corresponds to the model with
[Fe/H]$=-0.1$ and [$\alpha$/Fe]=$+0.2$ and highlights the fact that
$\Delta \teff$ is not a simple function of [Z/H].

The excellent agreement in the {\it slope} of the $\Delta \teff$
vs. [O/Fe]$+$[Fe/H] relation between the theoretical prediction and
the data suggests that we are reliably measuring the distribution of
effective temperatures for the stars in these galaxies (i.e., we are
solving for the detailed shape of the isochrones).  The flattening of
the observed trend at high $\sigma$ is interesting and will be
explored in detail in future work.  The offset between the two
relations is $\approx40$K, which is well within both the systematic
uncertainties in the color-temperature relations used to assign
temperatures to empirical stars and the theoretical uncertainties in
modeling RGB temperatures.  In order to provide a more direct
comparison between the observations and the theoretical predictions we
will need to create isochrones with the exact same chemical
composition as we derive for the early-type galaxies (e.g., variation
in C and N in addition to the $\alpha$ elements).  This work is
ongoing.

We remind the reader that a multitude of parameters are varying
simultaneously across the sample, and so it is impossible at present
to cleanly separate correlation from causation.  The correlation seen
in Figure \ref{fig:teff} could be driven by a third variable, and the
possibility remains that the $\Delta \teff$ parameter is soaking up an
unidentified model deficiency that varies with $\sigma$ and hence
[Z/H].

\subsection{Exploration of Systematic Uncertainties}
\label{s:sys}

\begin{figure}[!t]
\center
\resizebox{3.5in}{!}{\includegraphics{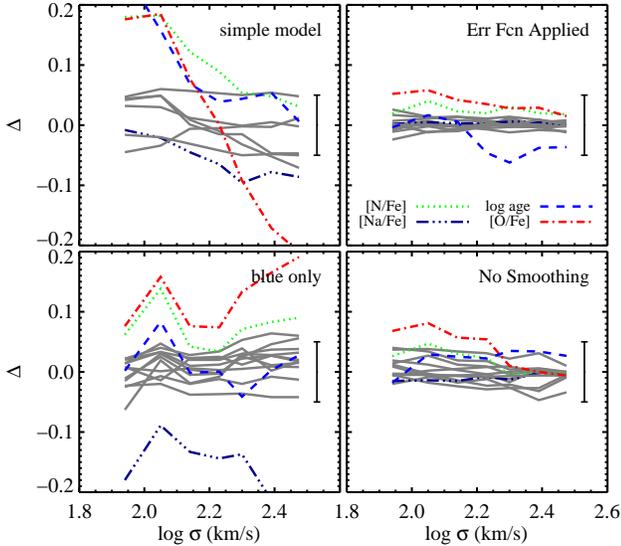}}
\caption{Effect on derived parameters due to varying one or more
  aspects of our analysis (plotted here as fiducial model minus
  modified model).  All parameters in Table \ref{t:res} are plotted
  here (except in the top left panel), but only the most deviant
  parameters are highlighted.  {\it Top Left Panel:} Variation between
  the standard model and a simple model that contains many fewer
  parameters. {\it Top Right Panel:} Variation between the standard
  model and one in which the mean residuals (averaged over all
  $\sigma$ bins) are added back into the model.  {\it Bottom Left
    Panel:} Variation in parameters when fitting only the blue
  spectral region ($\lambda<5800$\AA).  {\it Bottom Right Panel:}
  Variation in parameters when the spectra are not smoothed to a
  common dispersion of $\sigma=350\kms$.  The single error bar in each
  panel represents an error of $\pm0.05$ dex, a value we regard as a
  plausible upper limit to the systematic errors on the derived
  parameters.}
\label{fig:sys}
\end{figure}

In this section we explore variations in the fiducial model and in our
primary set of stacked spectra in order to quantify systematic
uncertainties in our main results.  In Figure \ref{fig:sys} we plot
the difference in either abundance ratio or log(age) between our main
results and a modification to our standard analysis.  In the upper
left panel we compare our standard analysis to a simplified model with
fewer free parameters.  In particular, in this model the IMF is fixed
to the \citet{Kroupa01} form, only a single age component is included
(rather than the standard two components), the iron peak elements are
forced to track Fe, and all the nuisance parameters are turned off
(including $\Delta \teff$).  This model results in a larger rms
residual between data and model compared to our fiducial model (i.e.,
it is a poorer fit), which should not be surprising given that there
are fewer free parameters.  Nonetheless, the resulting abundance
ratios agree well with the standard model, in particular Mg, C, Ca,
Si, Ti, and Fe agree to within 0.1 dex and in most cases to within
0.05 dex.  The largest outliers are the age at low $\sigma$ and the
[O/Fe] and [N/Fe] abundances.  The difference in [O/Fe] is
particularly strong.  We speculate that the larger range in [O/Fe] in
the simple model is due to the fact that the parameter $\Delta \teff$
is not included.  In essence, [O/Fe] is forced to vary more in order
to compensate for the lack of variation in $\Delta \teff$.

In the upper right panel we consider a model where the mean
wavelength-dependent residuals averaged over the sample are added back
into the model.  In other words, in this case the model is multiplied
by the mean residuals shown in Figure \ref{fig:resall} before the
model is fit to the data.  This will obviously result in a much lower
rms difference between the best-fit model and data, and the question
is whether or not this changes the derived parameters.  As
demonstrated in Figure \ref{fig:sys}, the resulting change in
abundance patterns is very small, typically $<0.03$ dex.  This implies
that the small mis-match between the data and our model does not bias
the resulting best-fit parameters.

In the lower left panel we consider the change in parameters when only
the blue spectral region ($\lambda<5800$\AA) is included in the fit.
Here again most parameters change very little, by $<0.05$ dex.
Significant outliers include [Na/Fe] and [O/Fe].  The change in
[Na/Fe] is not surprising because the two features most sensitive to
Na are the \ionn{Na}{i} features at 5895\AA\, and 8190\AA.  With these
features masked out the constraint on the Na abundance is coming
exclusively from the effect of Na on the electron pressure in the
stellar atmospheres, which indirectly affects the ionization states of
other elements.  For [O/Fe] the deviation is also significant,
exceeding 0.1 dex in several bins.  The origin of this offset is not
clear, although we note that the best-fit $\Delta \teff$ parameter is
lower in the case of fitting only the blue spectra.  With cooler stars
the metal-lines are stronger, perhaps alleviating the need for higher
[O/Fe] values.  This highlights the need for spectra that extend into
the red, where the $\Delta \teff$ parameter has a large effect and
hence can be more robustly measured.  Moreover, at $\lambda<5800$\AA\,
the constraint on [O/Fe] is coming primarily through the effect of O
on molecular dissociation equilibrium in the stellar atmospheres,
while at $\lambda>5800$\AA\, the effect of O on the spectrum is
primarily to increase the strength of the TiO lines (see Figures
\ref{fig:blue} and \ref{fig:red}).  The red spectrum therefore offers
at the very least a complementary, and arguably a stronger constraint
on the [O/Fe] abundance.

In the lower right panel we show the results of fitting to the
unsmoothed stacked spectra.  This is primarily a test of the model in
the low $\sigma$ bins, where the unsmoothed spectra have a resolution
$\approx4\times$ higher than the smoothed spectra.  Here again
essentially all parameters agree with our standard results to better
than 0.03 dex.  The largest outlier at low $\sigma$ is [O/Fe], but
even in this case the disagreement is $<0.1$ dex.

We have performed an additional test (not shown in the figure), where
we masked wavelengths bluer than 4200\AA.  This was done primarily to
test the [Co/Fe] result, since there are several strong \ionn{Co}{i}
lines with hyperfine splitting components at $\lambda<4200$\AA\, and
many weaker blends at longer wavelengths (see Figure \ref{fig:blue}).
The differences in the derived [Co/Fe] ratios between this and our
fiducial model is $<0.03$ dex for all $\sigma$ bins except for the
lowest bin, where the difference is 0.046 dex.  The result for Co
seems be insensitive to the particular wavelength range used in the
fit.

In summary, the tests performed in this section suggest that the
systematic uncertainties in our derived parameters is probably $<0.05$
dex. 

\begin{figure}
\center
\resizebox{3.5in}{!}{\includegraphics{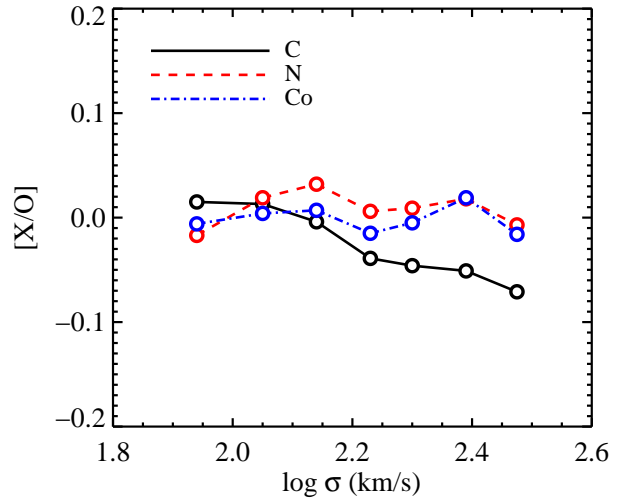}}
\caption{Abundance ratios between C, N, Co, and O as a function of
  early-type galaxy velocity dispersion.  All of these elements seem
  to track each other very closely, especially N, O, and Co.
  Systematic errors are probably $<0.05$ dex (see Section
  \ref{s:sys}).}
\label{fig:cnco}
\end{figure}

\section{Comparison to Previous Work}
\label{s:comp}

\begin{figure*}[!t]
\center
\resizebox{7in}{!}{\includegraphics{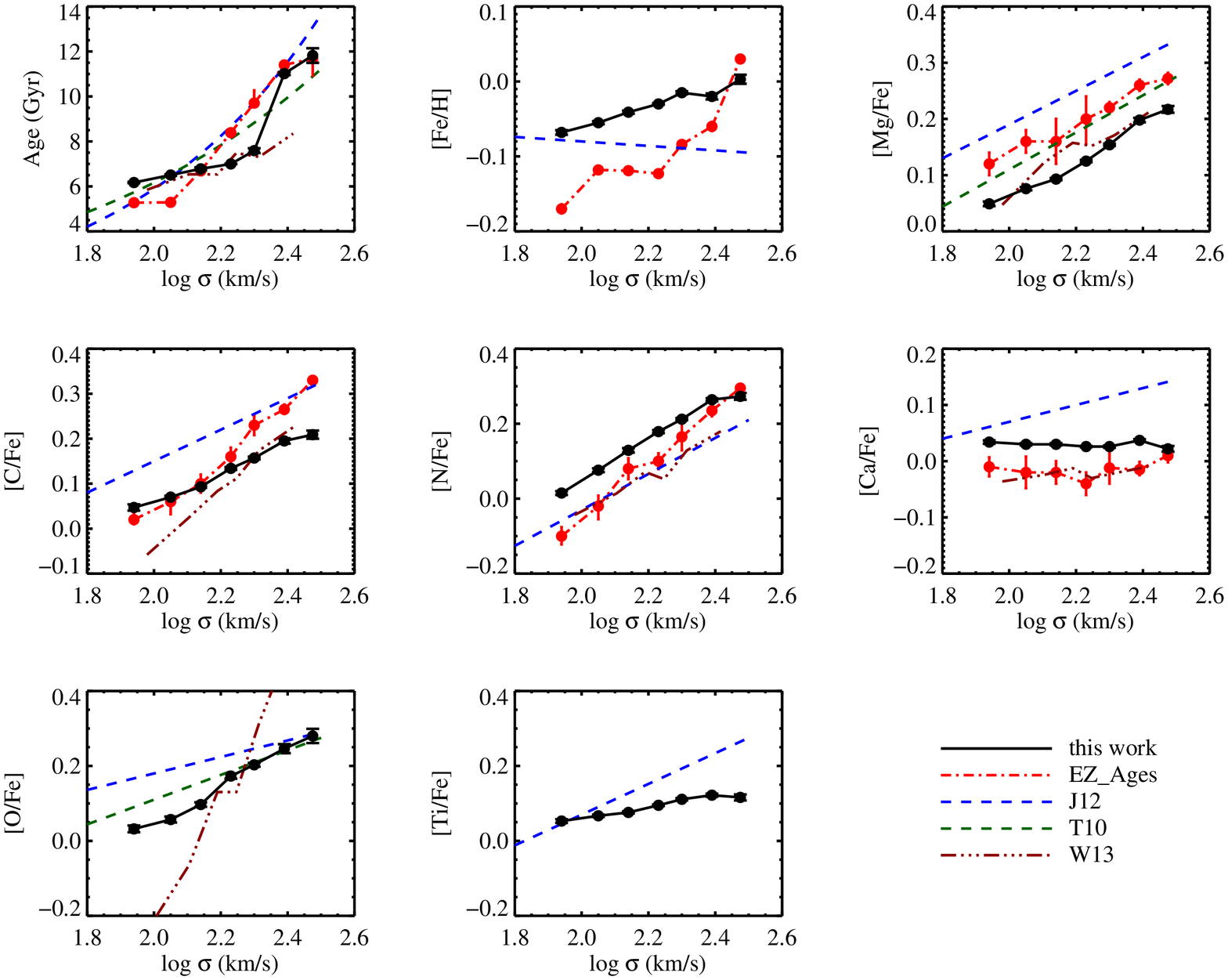}}
\caption{Comparison between different techniques for estimating ages
  and elemental abundances.  Our results are compared to techniques
  that rely on Lick index fitting, including
  \citet[][\texttt{EZ\_Ages}]{Graves08}, \citet[][T10]{Thomas10},
  \citet[][J12]{Johansson12}, and \citet[][W13]{Worthey13}.  The
  \texttt{EZ\_Ages} results are based on modeling the exact same
  stacked spectra as used herein, the W13 results are based on similar
  but not identical stacks, while the T10 and J12 results are linear
  fits to results obtained for $\sim4000$ SDSS early-type galaxies
  with light-weighted ages greater than $2.5$ Gyr.  Note that T10
  derive a single [$\alpha$/Fe] ratio, which we plot in both the
  [Mg/Fe] and [O/Fe] panels, as it is not entirely clear which element
  their [$\alpha$/Fe] parameter is most sensitive to.  Qualitatively
  the derived trends agree well between various methods/groups,
  although quantitative differences are clearly evident.}
\label{fig:lick}
\end{figure*}

In this section we provide a comparison to other techniques for
deriving stellar population parameters of early-type galaxies.  In
Figure \ref{fig:lick} we show abundance ratios and light-weighted ages
from several groups, including the results presented here, from the
\texttt{EZ\_Ages} code \citep{Graves08}, and from
\citet[][T10]{Thomas10}, \citet[][J12]{Johansson12}, and
\citet[][W13]{Worthey13}.

The \texttt{EZ\_Ages} code fits 7 Lick indices with the SPS models of
\citet{Schiavon07} in order to derive the age, [Fe/H], [C/Fe], [N/Fe],
[Mg/Fe], and [Ca/Fe].  The code was run on the exact same stacked
spectra analyzed herein, so that any differences can be attributed
entirely to differences in modeling techniques.  In general the
agreement is very good.  The [C/Fe] and [N/Fe] trends are stronger
with \texttt{EZ\_Ages} compared to the results presented heren.
Perhaps the most significant differences are the overall higher
[Mg/Fe] abundances in \texttt{EZ\_Ages} (by $\approx0.07$ dex), and
the lower [Fe/H] abundances (by $\approx0.07$ dex).

The results from T10 are based on the SPS models of \citet{Thomas03}
and \citet{Thomas04}.  T10 analyzed a morphologically-selected sample
of early-type galaxies from SDSS.  They used 25 Lick indices in order
to derive three parameters: [Z/H], age, and [$\alpha$/Fe].  We compare
their derived [$\alpha$/Fe] values to both [Mg/Fe] and [O/Fe].  We
find overall satisfactory agreement between their [$\alpha$/Fe] values
and our [Mg/Fe] and [O/Fe] values, and also between their ages and
ours.

J12 present results from the exact same morphologically-selected
sample of early-type galaxies analyzed in T10, but with the updated
SPS models from \citet{Thomas11}.  These models allow for variation in
the elements C, N, O, Mg, Ca, and Ti, in addition to metallicity and
age.  They fit their models to the data using 18 Lick indices.  The
differences between J12 and our own work is the largest of any of the
techniques shown in Figure \ref{fig:lick} (except for [O/Fe] from
W13), although the differences rarely exceed 0.1 dex.  It is difficult
to know if the differences are due to modeling techniques or sample
selection, especially at low $\sigma$ where there is likely to be
significant scatter between morphological (`early-type') and emission
line (`quiescent') selections.

Very recently, W13 analyzed stacked SDSS early-type galaxy spectra
using their SPS models.  The stacks are not identical to those
analyzed herein, but are very similar and so a direct comparison
between W13 and our results is possible.  From Figure \ref{fig:lick}
it is clear that the ages, [Mg/Fe], [N/Fe], and [Ca/Fe] trends are in
good agreement between the techniques, except for the age in the
highest $\sigma$ bin in W13.  The [C/Fe] abundances are also in
reasonable agreement.  In contrast, the [O/Fe] trend derived by W13 is
much steeper than not only our results but also those of J12 (and T10
if one treats their [$\alpha$/Fe] as [O/Fe]).  In addition, their
inferred [Fe/H] abundances\footnote{W13 (in addition to T10 and J12)
  do not constrain [Fe/H] directly but instead estimate [Z/H] and the
  abundance ratios.  [Fe/H] is thus a derived product and can be
  roughly estimated via [Fe/H]$=$[Z/H]$-A$\,[$\alpha$/Fe] with $A$ in
  the range $0.7-0.95$ depending on the detailed abundance pattern
  \citep{Trager00a}.  For the purposes of this discussion we estimated
  [Fe/H] from W13's data via [Fe/H]$=$[Z/H]$-0.9$[O/Fe].} range from
$-0.05$ at low $\sigma$ to $-0.7$ at high $\sigma$.  The results for
[Fe/H] are so different from the other results in Figure
\ref{fig:lick} that they were omitted from the Figure for clarity.
W13 fixed [Ti/Fe]$=0.0$ throughout their analysis.  These authors also
presented results for Si, finding [Si/Fe] is approximately 0.0 at low
$\sigma$ and increases to 0.5 at high $\sigma$.  This is also in
contrast to our results, in which [Si/Fe] increases from $0.0$ to
$0.16$ from low to high $\sigma$.  At present we do not know the
origins of these discrepancies.

\begin{figure*}[!t]
\center
\vspace{0.7cm}
\resizebox{7in}{!}{\includegraphics{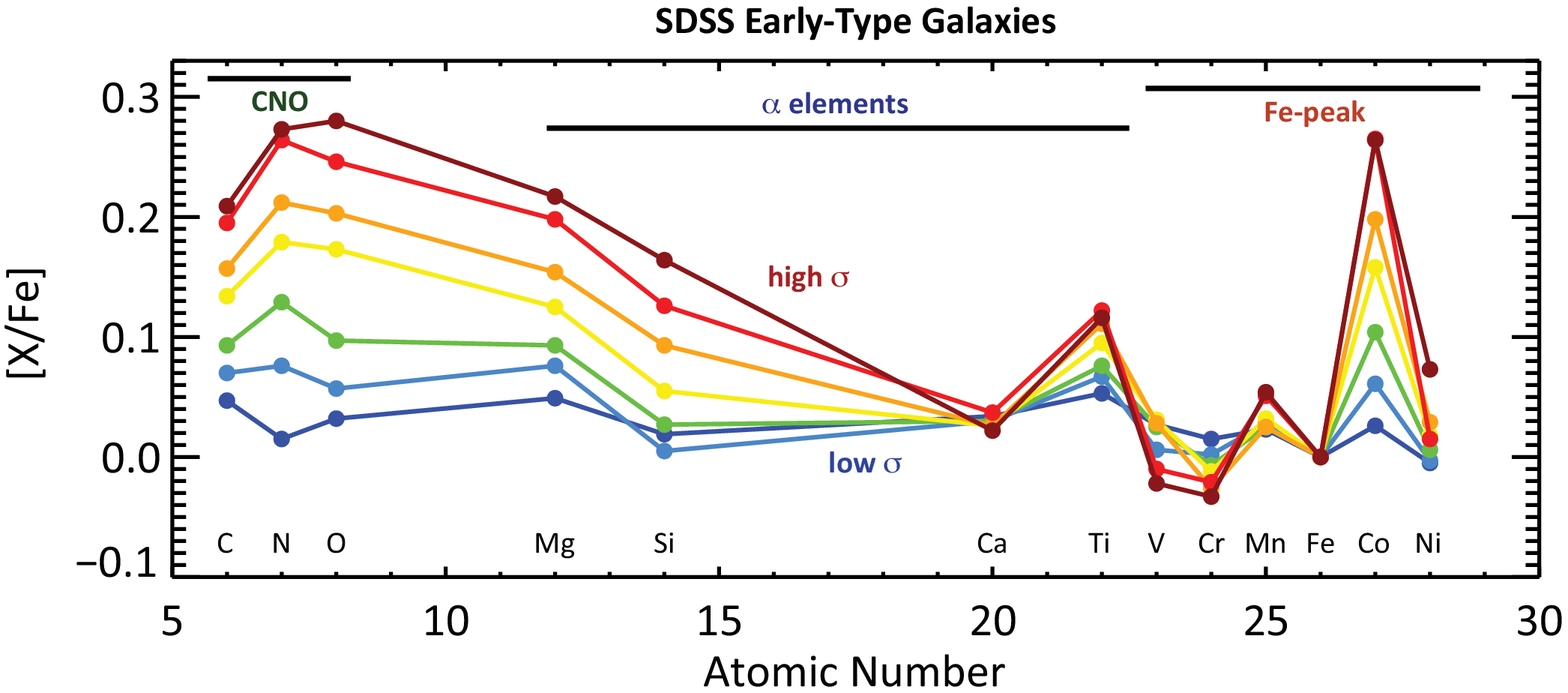}}
\caption{Summary of the abundance trends derived herein from stacked
  spectra of SDSS early-type galaxies.  The sample spans a range in
  velocity dispersion from $\sigma=90\kms$ to $\sigma=300\kms$.  The
  results for Na are omitted from this summary diagram, though we note
  that [Na/Fe] varies from $-0.16$ to $+0.43$ from low to high
  $\sigma$.}
\label{fig:sum}
\end{figure*}

One of the more notable features of Figure \ref{fig:lick} is the
[Mg/Fe] vs. $\sigma$ panel, where there is a $\sim0.1$ dex variation
in the normalization of the relation.  This variation is significant
in the context of using [Mg/Fe] to infer a star formation timescale
\citep[e.g,.][]{Thomas05}.  \citet{Trager08} noted that their models
produced [Mg/Fe] values lower than previous Lick index-based models,
and they speculated that this may be due to the different treatment of
the response functions.  In nearly all previous work on Lick index
modeling, three theoretical stellar spectra were combined to estimate
the response of the integrated light spectrum to abundance variations
\citep{Tripicco95, Korn05}.  Trager et al., and also
\citet{Worthey13}, instead used many stellar spectra covering the
isochrone in order to estimate response functions, in a manner similar
to what is done in the CvD models.  In addition, previous Lick
index-based models calculated response functions for an isochrone with
an age of 5 Gyr, whereas more recent models use older isochrones (or a
range of isochrones appropriate for the ages of the models).

We have estimated the effects of these assumptions by modifying our
model to employ only 3 stars (as opposed to the usual 20) to estimate
response functions, and also to use stars along a 5 Gyr rather than a
13 Gyr isochrone.  The number of stars used to compute response
functions has a significant effect.  We have created response
functions for the Mg {\it b} Lick index due to a factor of two change
in [$\alpha$/H].  A model using 20 stars has a response function in
the Mg {\it b} equivalent width (EW) that is 30\% greater than a model
that uses only 3 stars.  The effect of the age is rather modest, with
a change in EW of only 5\%, with the 13 Gyr isochrone producing
greater EWs.  The combination of both effects is a 37\% increase in
the response function of the Mg {\it b} index when 20 stars are used
along a 13 Gyr isochrone, compared to only 3 stars along a 5 Gyr
isochrone.  This translates directly into differences in the derived
[Mg/H] abundances, and therefore any model using only a small number
of stars along an intermediate-age isochrone will tend to
overestimate the [Mg/H] abundance.

In summary, there is broad agreement between models for those elements
that impart strong, localized changes in low resolution data
(especially C, N, Mg, and Ca; see Figure \ref{fig:blue}).  In such
cases models based on spectral indices seem to perform as well as our
model based on full spectrum fitting.  The power of fitting the
detailed spectrum lies in measuring parameters that either impart a
more subtle change in the spectrum (as in the case of Si and Mn), or
affect large swaths of the spectrum (as in the case of O, Ti, and most
of the iron peak elements).

\section{Discussion}
\label{s:disc}

The main results from this paper are summarized in Figure
\ref{fig:sum}, where we show the abundance patterns of 13 elements as
a function of early-type galaxy velocity dispersion.  This is the most
comprehensive analysis of the detailed abundance patterns of the stars
within distant galaxies to-date.  We emphasize that the spectra sample
approximately the inner $0.5R_e$, and thus the abundances derived
should be representative of the inner regions of the galaxies. In this
section we discuss several implications of these results.

The qualitative agreement between our results and those of previous
authors is, for the most part, very encouraging.  This is significant
because our model is a major departure from all previous analyses of
the abundance patterns of early-type galaxies.  Previous work focused
on the analysis of selected spectral indices with theoretical spectral
models that were relatively simple with respect to the theoretical
spectral libraries used herein.  In addition, we have included a large
number of additional free parameters in order to marginalize over our
ignorance of various aspects of stellar evolution and stellar
populations.  In spite of these differences in techniques, there is
excellent agrement in the trends, including the derived light-weighted
ages, the weak variation in [Fe/H] and [Ca/Fe], and the strong
variation in [Mg/Fe], [C/Fe], and [N/Fe].  It should not be surprising
that there is good agreement between techniques for the elements C, N,
Mg, and Ca because these elements impart strong, localized changes in
low resolution spectra (see Figure \ref{fig:blue}).  Quantitatively
there are differences in the slope and normalization for some of these
relations, though the differences rarely exceed 0.1 dex in amplitude.
Based on our preliminary investigations, it appears that at least some
of the differences can be traced back to the number of theoretical
spectra used to estimate response functions to elemental abundance
variations \citep[see also][]{Trager08}.  The real power of fitting
the detailed spectrum is in the measurement of parameters that either
impart a small change in the spectrum (e.g., Mn, Sr, Ba), or that
affect large regions of the spectrum (e.g., O, Ti, V, Cr, Mn, Fe, Co,
Ni).  For an example of the former, see \citet{Conroy13a}.

The standard interpretation of [$\alpha$/Fe] is that this ratio is
sensitive to the time-scale of star formation (SF), with higher values
corresponding to shorter timescales \citep[e.g.,][]{Tinsley79,
  Thomas99}.  As mentioned in the Introduction, this ratio is also
sensitive to the IMF, the nucleosynthetic yields, the delay time
distribution of Type Ia SNe, and the preferential loss of metals via
winds.  Some chemical evolution models are able to reproduce the
observed variation in not only Mg and O but also Si and Ca based on
the assumption that the main variable driving these trends is indeed
the SF timescale \citep{Pipino09}, while others also require variation
in other parameters such as the IMF \citep{Arrigoni10}.  In the
context of any chemical evolution model, the [Mg/Fe] ratio will be
correlated with the SF timescale, and so the lower ratios that we
derive herein have important implications for the inferred formation
timescales of the stars within these massive galaxies.  As an example
of the effect, \citet{Thomas05} use the equation [$\alpha$/Fe]$\approx
\frac{1}{5}-\frac{1}{6}\,{\rm log}\,\Delta t$, calibrated from the
simple chemical evolution model of \citet{Thomas99}, in order to
convert their measured [$\alpha$/Fe] ratios into constraints on the
formation histories of early-type galaxies.  In our analysis we find
[Mg/Fe]$=0.22$ in the most massive galaxies, which, when employing the
equation above, corresponds to a timescale of 0.8 Gyr.  In contrast,
the \texttt{EZ\_Ages} model and T10 find [Mg/Fe]$\approx0.27$ at the
highest dispersions, corresponding to a timescale of 0.4 Gyr.  Based
on the models of \citet{Thomas11}, J12 find [Mg/Fe]$\approx0.33$ at
the highest dispersions, corresponding to a remarkably short timescale
of 0.2 Gyr.  It is interesting in this context that the [O/Fe] ratios
derived by J12 are in much better agreement with our results at the
highest dispersions, and so if [O/Fe] were used as a tracer of
$\alpha$, rather than Mg, then the inferred timescales would be more
similar between different groups at high $\sigma$.  However, this of
course implies that [O/Mg]$\ne0$, and in this case it is not clear
that a single parameter ``[$\alpha$/Fe]'' can be used to simply (or
reliably) translate derived abundance ratios into SF timescales.
Clearly sophisticated chemical evolution models are need to provide
further insight.

One of the more interesting trends evident in Figure \ref{fig:sum}
(see also Figure \ref{fig:cnco}) is the fact that the iron peak
element Co closely tracks O rather than Fe.  This is the first
measurement of Co in the integrated light spectra of early-type
galaxies, and it provides an important constraint on the
nucleosynthetic origin of Co.  Taken at face value,
[Co/Fe]$\approx$[O/Fe] implies that Co forms primarily in massive
stars, in apparent contrast to the other iron peak elements.  In this
context it is worth recalling that Co is also peculiar amongst the
iron peak elements in the metal-poor Galactic halo
\citep{McWilliam97}.  This may be a coincidence at some level, as the
abundance pattern of the very metal-poor halo is probably due
exclusively to very metal-poor Type II SNe, in contrast to the
processes operating to shape the abundance patterns in massive
early-type galaxies.  Core-collapse SNe nucleosynthesis models predict
that Co is synthesized by complete Si-burning in the deepest layers,
while Cr and Mn are produced in the outer incomplete Si-burning layers
\citep{Woosley95, Nakamura99}.  There is thus a mechanism by which to
separate the production of Co from the other iron peak elements.
Perhaps more significantly, the Type Ia SNe yields from
\citet{Nomoto84} show a deficit of Co production compared to other
iron peak elements.  Again, detailed chemical evolution models will be
required to sort out these details.

We reaffirm previous results that the light $\alpha$ elements (e.g., O
and Mg) show greater variation than the heavier $\alpha$ elements
(e.g., Ca and Ti).  We also find in this work that the variation in Si
is intermediate between the lighter and heavier elements.  At least
some of these trends can be understood in the context of Type Ia SNe
yields.  The \citet{Nomoto84} models produce a large quantity of Ca
(comparable to Cr, Mn, Fe, and Ni), and much more Si than either O or
Mg.  The modest variation in Ti is somewhat more difficult to
understand in this context, because the Nomoto et al. models produce
very little Ti (comparable to O and Mg).  Whether or not the variation
in [Ti/Fe] with galaxy dispersion can be reproduced with standard
nucleosynthesis yield tables remains to be seen.

Finally we turn to C and N.  As found in previous work, both [C/Fe]
and [N/Fe] increase with increasing dispersion \citep{Schiavon07,
  Graves07, Smith09, Johansson12}.  We find that both of these
elements track O and Mg quite closely (in detail C seems to track Mg
while N tracks O, but we regard this level of similarity as
fortuitous).  \citet{Johansson12} also find that C closely tracks O,
with [C/O]$\sim0.0$, but they find sub-solar [N/O] values.  The fact
that C and N are tracking the light $\alpha$ elements is probably
telling us either that C and N form mostly in massive stars, or that
there is a lower limit to the star formation timescale such that both
massive stars and moderately massive AGB stars (which can produce lots
of C and N) are always contributing to the enrichment of the ISM
\citep[see also][who come to similar conclusions]{Schiavon07,
  Johansson12}.  The latter interpretation is consistent with the star
formation timescales derived using our [Mg/Fe] ratios.  Much effort is
still needed on the theoretical side to understand the nucleosynthetic
origins of these elements in order to use them as cosmological clocks.

\section{Summary}
\label{s:sum}

In this paper we have used the CvD population synthesis model to fit
very high quality stacked spectra of SDSS early-type galaxies spanning
a range in velocity dispersion from $90\kms$ to $300\kms$.  The novel
feature of our approach is fitting the full continuum-normalized
optical-NIR spectra ($4000$\AA$-8800$\AA), rather than a select number
of indices.  The model includes variation in age, the abundances of 16
elements, the effective temperature distribution of the stars, the
stellar IMF, amongst other parameters.  The resulting quality of the
fits are very high, with rms residuals ranging from $0.2\%-0.3$\%
across the sample.  A variety of tests reveal that the systematic
uncertainties in our measurements are probably 0.05 dex or less.  We
now summarize our main results.

\begin{itemize}

\item Across the sample from low to high $\sigma$, ages increase from
  6 to 12 Gyr, [Fe/H] increases from $-0.07$ to 0.00, and the
  abundance ratios [O/Fe], [Na/Fe], [Mg/Fe], [Si/Fe], [Ti/Fe], and
  [Co/Fe] increase from approximately zero to $0.1-0.4$, with O, Na,
  Mg reaching the highest values.  In contrast, [Ca/Fe], [V/Fe],
  [Cr/Fe], [Mn/Fe], and [Ni/Fe] remain nearly constant as $\sigma$
  increases.

\item N and Co track O to better than 0.03 dex over the full sample,
  and C tracks O to better than 0.1 dex.  Co therefore probably forms
  predominantly in massive stars.  For N and C the situation is more
  complex because these elements probably form in both massive stars
  and AGB stars.  The constant values of [C/O] and [N/O] therefore
  either requires C and N to form mostly in massive stars or the star
  formation timescale to never be shorter than several hundred Myr,
  the lifetime of a $\sim3\,\Msun$ star.

\item Our model allows us to measure the shift in the effective
  temperature of the stars with respect to a solar metallicity
  isochrone.  This shift inversely correlates with the total
  metallicity, with a slope that is in excellent agreement with
  theoretical expectations.  We are therefore able to measure not only
  the mean ages and abundance patterns but also the distribution of
  effective temperatures of the stars in early-type galaxies.

\item Comparison of our results to Lick index-based techniques reveals
  agreement in several quantities, such as age, [N/Fe], and [Ca/Fe],
  and less agreement in others including [C/Fe], [Fe/H], and [Mg/Fe].
  Some of the disagreement may be due to different samples of
  early-type galaxies, but the different modeling techniques also
  seems to play an important role.  Perhaps most relevant from the
  standpoint of broader implications, we find lower [Mg/Fe] values at
  fixed $\sigma$ (by $0.05-0.1$ dex) than some previous work
  \citep[especially those which employ the][response tables]{Korn05},
  which we believe is due to our more accurate modeling of spectral
  variations due to abundance changes.  This implies longer formation
  timescales for massive early-type galaxies than previously reported.

\end{itemize}

These results will set the foundation for future studies aimed at
constraining the evolutionary histories of galaxies from their
detailed abundance patterns.  With such a large number of elements
precisely measured over a wide range in galaxy mass, it should be
possible to place novel constraints not only on the formation
histories of these galaxies but also on the nucleosynthetic origins of
elements that are difficult to model from first principles.


\acknowledgments 

We thank Bob Kurucz for his continued assistance with the line lists
and model atmospheres.  We also thank Judy Cohen, Jonas Johansson,
Andy McWilliam, Ricardo Schiavon, Scott Trager, Stan Woosley, and Guy
Worthey for useful discussions.  The referee, Ricardo Schiavon, is
thanked for a thorough and very helpful report.  This material is
based upon work supported by the National Science Foundation under
Grant Number 1229745.

Funding for the SDSS and SDSS-II has been provided by the Alfred
P. Sloan Foundation, the Participating Institutions, the National
Science Foundation, the U.S. Department of Energy, the National
Aeronautics and Space Administration, the Japanese Monbukagakusho, the
Max Planck Society, and the Higher Education Funding Council for
England. The SDSS Web Site is http://www.sdss.org/.  The SDSS is
managed by the Astrophysical Research Consortium for the Participating
Institutions. The Participating Institutions are the American Museum
of Natural History, Astrophysical Institute Potsdam, University of
Basel, University of Cambridge, Case Western Reserve University,
University of Chicago, Drexel University, Fermilab, the Institute for
Advanced Study, the Japan Participation Group, Johns Hopkins
University, the Joint Institute for Nuclear Astrophysics, the Kavli
Institute for Particle Astrophysics and Cosmology, the Korean
Scientist Group, the Chinese Academy of Sciences (LAMOST), Los Alamos
National Laboratory, the Max-Planck-Institute for Astronomy (MPIA),
the Max-Planck-Institute for Astrophysics (MPA), New Mexico State
University, Ohio State University, University of Pittsburgh,
University of Portsmouth, Princeton University, the United States
Naval Observatory, and the University of Washington.


\begin{thebibliography}{73}
\expandafter\ifx\csname natexlab\endcsname\relax\def\natexlab#1{#1}\fi

\bibitem[{{Abazajian} {et~al.}(2009){Abazajian}, {Adelman-McCarthy},
  {Ag{\"u}eros}, {Allam}, {Allende Prieto}, {An}, {Anderson}, {Anderson},
  {Annis}, {Bahcall}, \& et~al.}]{abazajian09}
{Abazajian}, K.~N., {Adelman-McCarthy}, J.~K., {Ag{\"u}eros}, M.~A., {Allam},
  S.~S., {Allende Prieto}, C., {An}, D., {Anderson}, K.~S.~J., {Anderson},
  S.~F., {Annis}, J., {Bahcall}, N.~A., \& et~al. 2009, \apjs, 182, 543

\bibitem[{{Arrigoni} {et~al.}(2010){Arrigoni}, {Trager}, {Somerville}, \&
  {Gibson}}]{Arrigoni10}
{Arrigoni}, M., {Trager}, S.~C., {Somerville}, R.~S., \& {Gibson}, B.~K. 2010,
  \mnras, 402, 173

\bibitem[{{Bochanski} {et~al.}(2007){Bochanski}, {West}, {Hawley}, \&
  {Covey}}]{Bochanski07}
{Bochanski}, J.~J., {West}, A.~A., {Hawley}, S.~L., \& {Covey}, K.~R. 2007,
  \aj, 133, 531

\bibitem[{{Burstein} {et~al.}(1984){Burstein}, {Faber}, {Gaskell}, \&
  {Krumm}}]{Burstein84}
{Burstein}, D., {Faber}, S.~M., {Gaskell}, C.~M., \& {Krumm}, N. 1984, \apj,
  287, 586

\bibitem[{{Carretta} {et~al.}(2001){Carretta}, {Cohen}, {Gratton}, \&
  {Behr}}]{Carretta01}
{Carretta}, E., {Cohen}, J.~G., {Gratton}, R.~G., \& {Behr}, B.~B. 2001, \aj,
  122, 1469

\bibitem[{{Cenarro} {et~al.}(2003){Cenarro}, {Gorgas}, {Vazdekis}, {Cardiel},
  \& {Peletier}}]{Cenarro03}
{Cenarro}, A.~J., {Gorgas}, J., {Vazdekis}, A., {Cardiel}, N., \& {Peletier},
  R.~F. 2003, \mnras, 339, L12

\bibitem[{{Cid Fernandes} {et~al.}(2005){Cid Fernandes}, {Mateus}, {Sodr{\'e}},
  {Stasi{\'n}ska}, \& {Gomes}}]{CidFernandes05}
{Cid Fernandes}, R., {Mateus}, A., {Sodr{\'e}}, L., {Stasi{\'n}ska}, G., \&
  {Gomes}, J.~M. 2005, \mnras, 358, 363

\bibitem[{{Conroy}(2013)}]{Conroy13b}
{Conroy}, C. 2013, \araa, 51, 393

\bibitem[{{Conroy} \& {van Dokkum}(2012{\natexlab{a}})}]{Conroy12a}
{Conroy}, C. \& {van Dokkum}, P. 2012{\natexlab{a}}, \apj, 747, 69

\bibitem[{{Conroy} \& {van Dokkum}(2012{\natexlab{b}})}]{Conroy12b}
{Conroy}, C. \& {van Dokkum}, P.~G. 2012{\natexlab{b}}, \apj, 760, 71

\bibitem[{{Conroy} {et~al.}(2013){Conroy}, {van Dokkum}, \&
  {Graves}}]{Conroy13a}
{Conroy}, C., {van Dokkum}, P.~G., \& {Graves}, G.~J. 2013, \apjl, 763, L25

\bibitem[{{Cushing} {et~al.}(2005){Cushing}, {Rayner}, \& {Vacca}}]{Cushing05}
{Cushing}, M.~C., {Rayner}, J.~T., \& {Vacca}, W.~D. 2005, \apj, 623, 1115

\bibitem[{{Dotter} {et~al.}(2007){Dotter}, {Chaboyer}, {Ferguson}, {Lee},
  {Worthey}, {Jevremovi{\'c}}, \& {Baron}}]{Dotter07}
{Dotter}, A., {Chaboyer}, B., {Ferguson}, J.~W., {Lee}, H.-c., {Worthey}, G.,
  {Jevremovi{\'c}}, D., \& {Baron}, E. 2007, \apj, 666, 403

\bibitem[{{Dotter} {et~al.}(2008){Dotter}, {Chaboyer}, {Jevremovi{\'c}},
  {Kostov}, {Baron}, \& {Ferguson}}]{Dotter08b}
{Dotter}, A., {Chaboyer}, B., {Jevremovi{\'c}}, D., {Kostov}, V., {Baron}, E.,
  \& {Ferguson}, J.~W. 2008, \apjs, 178, 89

\bibitem[{{Feltzing} \& {Johnson}(2002)}]{Feltzing02}
{Feltzing}, S. \& {Johnson}, R.~A. 2002, \aap, 385, 67

\bibitem[{{Graves} {et~al.}(2010){Graves}, {Faber}, \& {Schiavon}}]{graves10}
{Graves}, G.~J., {Faber}, S.~M., \& {Schiavon}, R.~P. 2010, \apj, 721, 278

\bibitem[{{Graves} {et~al.}(2007){Graves}, {Faber}, {Schiavon}, \&
  {Yan}}]{Graves07}
{Graves}, G.~J., {Faber}, S.~M., {Schiavon}, R.~P., \& {Yan}, R. 2007, \apj,
  671, 243

\bibitem[{{Graves} \& {Schiavon}(2008)}]{Graves08}
{Graves}, G.~J. \& {Schiavon}, R.~P. 2008, \apjs, 177, 446

\bibitem[{{Heavens} {et~al.}(2000){Heavens}, {Jimenez}, \& {Lahav}}]{Heavens00}
{Heavens}, A.~F., {Jimenez}, R., \& {Lahav}, O. 2000, \mnras, 317, 965

\bibitem[{{Jimenez} {et~al.}(2007){Jimenez}, {Bernardi}, {Haiman}, {Panter}, \&
  {Heavens}}]{Jimenez07}
{Jimenez}, R., {Bernardi}, M., {Haiman}, Z., {Panter}, B., \& {Heavens}, A.~F.
  2007, \apj, 669, 947

\bibitem[{{Johansson} {et~al.}(2012){Johansson}, {Thomas}, \&
  {Maraston}}]{Johansson12}
{Johansson}, J., {Thomas}, D., \& {Maraston}, C. 2012, \mnras, 421, 1908

\bibitem[{{Kelson} {et~al.}(2006){Kelson}, {Illingworth}, {Franx}, \& {van
  Dokkum}}]{Kelson06}
{Kelson}, D.~D., {Illingworth}, G.~D., {Franx}, M., \& {van Dokkum}, P.~G.
  2006, \apj, 653, 159

\bibitem[{{Korn} {et~al.}(2005){Korn}, {Maraston}, \& {Thomas}}]{Korn05}
{Korn}, A.~J., {Maraston}, C., \& {Thomas}, D. 2005, \aap, 438, 685

\bibitem[{{Kroupa}(2001)}]{Kroupa01}
{Kroupa}, P. 2001, \mnras, 322, 231

\bibitem[{{Kurucz}(1970)}]{Kurucz70}
{Kurucz}, R.~L. 1970, SAO Special Report, 309

\bibitem[{{Kurucz}(1993)}]{Kurucz93}
---. 1993, {SYNTHE spectrum synthesis programs and line data}, ed. {Kurucz,
  R.~L.}

\bibitem[{{Magic} {et~al.}(2010){Magic}, {Serenelli}, {Weiss}, \&
  {Chaboyer}}]{Magic10}
{Magic}, Z., {Serenelli}, A., {Weiss}, A., \& {Chaboyer}, B. 2010, \apj, 718,
  1378

\bibitem[{{McWilliam}(1997)}]{McWilliam97}
{McWilliam}, A. 1997, \araa, 35, 503

\bibitem[{{Milone} {et~al.}(2000){Milone}, {Barbuy}, \& {Schiavon}}]{Milone00}
{Milone}, A., {Barbuy}, B., \& {Schiavon}, R.~P. 2000, \aj, 120, 131

\bibitem[{{Momany} {et~al.}(2003){Momany}, {Ortolani}, {Held}, {Barbuy},
  {Bica}, {Renzini}, {Bedin}, {Rich}, \& {Marconi}}]{Momany03}
{Momany}, Y., {Ortolani}, S., {Held}, E.~V., {Barbuy}, B., {Bica}, E.,
  {Renzini}, A., {Bedin}, L.~R., {Rich}, R.~M., \& {Marconi}, G. 2003, \aap,
  402, 607

\bibitem[{{Nakamura} {et~al.}(1999){Nakamura}, {Umeda}, {Nomoto}, {Thielemann},
  \& {Burrows}}]{Nakamura99}
{Nakamura}, T., {Umeda}, H., {Nomoto}, K., {Thielemann}, F.-K., \& {Burrows},
  A. 1999, \apj, 517, 193

\bibitem[{{Nomoto} {et~al.}(1984){Nomoto}, {Thielemann}, \& {Yokoi}}]{Nomoto84}
{Nomoto}, K., {Thielemann}, F.-K., \& {Yokoi}, K. 1984, \apj, 286, 644

\bibitem[{{Ocvirk} {et~al.}(2006){Ocvirk}, {Pichon}, {Lan{\c c}on}, \&
  {Thi{\'e}baut}}]{Ocvirk06}
{Ocvirk}, P., {Pichon}, C., {Lan{\c c}on}, A., \& {Thi{\'e}baut}, E. 2006,
  \mnras, 365, 46

\bibitem[{{Origlia} {et~al.}(2005){Origlia}, {Valenti}, \& {Rich}}]{Origlia05}
{Origlia}, L., {Valenti}, E., \& {Rich}, R.~M. 2005, \mnras, 356, 1276

\bibitem[{{Peek} \& {Graves}(2010)}]{peek10}
{Peek}, J.~E.~G. \& {Graves}, G.~J. 2010, \apj, 719, 415

\bibitem[{{Pipino} {et~al.}(2009){Pipino}, {Chiappini}, {Graves}, \&
  {Matteucci}}]{Pipino09}
{Pipino}, A., {Chiappini}, C., {Graves}, G., \& {Matteucci}, F. 2009, \mnras,
  396, 1151

\bibitem[{{Rayner} {et~al.}(2009){Rayner}, {Cushing}, \& {Vacca}}]{Rayner09}
{Rayner}, J.~T., {Cushing}, M.~C., \& {Vacca}, W.~D. 2009, \apjs, 185, 289

\bibitem[{{Saglia} {et~al.}(2002){Saglia}, {Maraston}, {Thomas}, {Bender}, \&
  {Colless}}]{Saglia02}
{Saglia}, R.~P., {Maraston}, C., {Thomas}, D., {Bender}, R., \& {Colless}, M.
  2002, \apjl, 579, L13

\bibitem[{{Salaris} {et~al.}(1993){Salaris}, {Chieffi}, \&
  {Straniero}}]{Salaris93}
{Salaris}, M., {Chieffi}, A., \& {Straniero}, O. 1993, \apj, 414, 580

\bibitem[{{Salpeter}(1955)}]{Salpeter55}
{Salpeter}, E.~E. 1955, \apj, 121, 161

\bibitem[{{S{\'a}nchez-Bl{\'a}zquez} {et~al.}(2006){S{\'a}nchez-Bl{\'a}zquez},
  {Peletier}, {Jim{\'e}nez-Vicente}, {Cardiel}, {Cenarro},
  {Falc{\'o}n-Barroso}, {Gorgas}, {Selam}, \& {Vazdekis}}]{Sanchez-Blazquez06}
{S{\'a}nchez-Bl{\'a}zquez}, P., {Peletier}, R.~F., {Jim{\'e}nez-Vicente}, J.,
  {Cardiel}, N., {Cenarro}, A.~J., {Falc{\'o}n-Barroso}, J., {Gorgas}, J.,
  {Selam}, S., \& {Vazdekis}, A. 2006, \mnras, 371, 703

\bibitem[{{Sbordone} {et~al.}(2004){Sbordone}, {Bonifacio}, {Castelli}, \&
  {Kurucz}}]{Sbordone04}
{Sbordone}, L., {Bonifacio}, P., {Castelli}, F., \& {Kurucz}, R.~L. 2004,
  Memorie della Societa Astronomica Italiana Supplementi, 5, 93

\bibitem[{{Schiavon}(2007)}]{Schiavon07}
{Schiavon}, R.~P. 2007, \apjs, 171, 146

\bibitem[{{Schiavon} {et~al.}(2004){Schiavon}, {Caldwell}, \&
  {Rose}}]{Schiavon04}
{Schiavon}, R.~P., {Caldwell}, N., \& {Rose}, J.~A. 2004, \aj, 127, 1513

\bibitem[{{Schiavon} {et~al.}(2005){Schiavon}, {Rose}, {Courteau}, \&
  {MacArthur}}]{Schiavon05}
{Schiavon}, R.~P., {Rose}, J.~A., {Courteau}, S., \& {MacArthur}, L.~A. 2005,
  \apjs, 160, 163

\bibitem[{{Shetrone} \& {Sandquist}(2000)}]{Shetrone00}
{Shetrone}, M.~D. \& {Sandquist}, E.~L. 2000, \aj, 120, 1913

\bibitem[{{Smith} {et~al.}(2009){Smith}, {Lucey}, {Hudson}, \&
  {Bridges}}]{Smith09}
{Smith}, R.~J., {Lucey}, J.~R., {Hudson}, M.~J., \& {Bridges}, T.~J. 2009,
  \mnras, 398, 119

\bibitem[{{Springob} {et~al.}(2012){Springob}, {Magoulas}, {Proctor},
  {Colless}, {Jones}, {Kobayashi}, {Campbell}, {Lucey}, \&
  {Mould}}]{springob12}
{Springob}, C.~M., {Magoulas}, C., {Proctor}, R., {Colless}, M., {Jones},
  D.~H., {Kobayashi}, C., {Campbell}, L., {Lucey}, J., \& {Mould}, J. 2012,
  \mnras, 420, 2773

\bibitem[{{Strauss} {et~al.}(2002)}]{strauss02}
{Strauss}, M.~A. {et~al.} 2002, \aj, 124, 1810

\bibitem[{{Tautvai{\v s}iene} {et~al.}(2000){Tautvai{\v s}iene}, {Edvardsson},
  {Tuominen}, \& {Ilyin}}]{Tautvaisiene00}
{Tautvai{\v s}iene}, G., {Edvardsson}, B., {Tuominen}, I., \& {Ilyin}, I. 2000,
  \aap, 360, 499

\bibitem[{{Thomas} {et~al.}(1999){Thomas}, {Greggio}, \& {Bender}}]{Thomas99}
{Thomas}, D., {Greggio}, L., \& {Bender}, R. 1999, \mnras, 302, 537

\bibitem[{{Thomas} {et~al.}(2003{\natexlab{a}}){Thomas}, {Maraston}, \&
  {Bender}}]{Thomas03b}
{Thomas}, D., {Maraston}, C., \& {Bender}, R. 2003{\natexlab{a}}, \mnras, 343,
  279

\bibitem[{{Thomas} {et~al.}(2003{\natexlab{b}}){Thomas}, {Maraston}, \&
  {Bender}}]{Thomas03}
---. 2003{\natexlab{b}}, \mnras, 339, 897

\bibitem[{{Thomas} {et~al.}(2005){Thomas}, {Maraston}, {Bender}, \& {Mendes de
  Oliveira}}]{Thomas05}
{Thomas}, D., {Maraston}, C., {Bender}, R., \& {Mendes de Oliveira}, C. 2005,
  \apj, 621, 673

\bibitem[{{Thomas} {et~al.}(2011){Thomas}, {Maraston}, \&
  {Johansson}}]{Thomas11}
{Thomas}, D., {Maraston}, C., \& {Johansson}, J. 2011, \mnras, 412, 2183

\bibitem[{{Thomas} {et~al.}(2004){Thomas}, {Maraston}, \& {Korn}}]{Thomas04}
{Thomas}, D., {Maraston}, C., \& {Korn}, A. 2004, \mnras, 351, L19

\bibitem[{{Thomas} {et~al.}(2010){Thomas}, {Maraston}, {Schawinski}, {Sarzi},
  \& {Silk}}]{Thomas10}
{Thomas}, D., {Maraston}, C., {Schawinski}, K., {Sarzi}, M., \& {Silk}, J.
  2010, \mnras, 404, 1775

\bibitem[{{Tinsley}(1979)}]{Tinsley79}
{Tinsley}, B.~M. 1979, \apj, 229, 1046

\bibitem[{{Tojeiro} {et~al.}(2009){Tojeiro}, {Wilkins}, {Heavens}, {Panter}, \&
  {Jimenez}}]{Tojeiro09}
{Tojeiro}, R., {Wilkins}, S., {Heavens}, A.~F., {Panter}, B., \& {Jimenez}, R.
  2009, \apjs, 185, 1

\bibitem[{{Trager} {et~al.}(2008){Trager}, {Faber}, \& {Dressler}}]{Trager08}
{Trager}, S.~C., {Faber}, S.~M., \& {Dressler}, A. 2008, \mnras, 386, 715

\bibitem[{{Trager} {et~al.}(2000){Trager}, {Faber}, {Worthey}, \&
  {Gonz{\'a}lez}}]{Trager00a}
{Trager}, S.~C., {Faber}, S.~M., {Worthey}, G., \& {Gonz{\'a}lez}, J.~J. 2000,
  \aj, 119, 1645

\bibitem[{{Trager} {et~al.}(1998){Trager}, {Worthey}, {Faber}, {Burstein}, \&
  {Gonzalez}}]{Trager98}
{Trager}, S.~C., {Worthey}, G., {Faber}, S.~M., {Burstein}, D., \& {Gonzalez},
  J.~J. 1998, \apjs, 116, 1

\bibitem[{{Tripicco} \& {Bell}(1995)}]{Tripicco95}
{Tripicco}, M.~J. \& {Bell}, R.~A. 1995, \aj, 110, 3035

\bibitem[{{van Dokkum} \& {Conroy}(2012)}]{vanDokkum12}
{van Dokkum}, P.~G. \& {Conroy}, C. 2012, \apj, 760, 70

\bibitem[{{VandenBerg} {et~al.}(2012){VandenBerg}, {Bergbusch}, {Dotter},
  {Ferguson}, {Michaud}, {Richer}, \& {Proffitt}}]{VandenBerg12}
{VandenBerg}, D.~A., {Bergbusch}, P.~A., {Dotter}, A., {Ferguson}, J.~W.,
  {Michaud}, G., {Richer}, J., \& {Proffitt}, C.~R. 2012, \apj, 755, 15

\bibitem[{{VandenBerg} {et~al.}(2007){VandenBerg}, {Gustafsson}, {Edvardsson},
  {Eriksson}, \& {Ferguson}}]{vandenberg07}
{VandenBerg}, D.~A., {Gustafsson}, B., {Edvardsson}, B., {Eriksson}, K., \&
  {Ferguson}, J. 2007, \apjl, 666, L105

\bibitem[{{Walcher} {et~al.}(2009){Walcher}, {Coelho}, {Gallazzi}, \&
  {Charlot}}]{Walcher09}
{Walcher}, C.~J., {Coelho}, P., {Gallazzi}, A., \& {Charlot}, S. 2009, \mnras,
  398, L44

\bibitem[{{Woosley} \& {Weaver}(1995)}]{Woosley95}
{Woosley}, S.~E. \& {Weaver}, T.~A. 1995, \apjs, 101, 181

\bibitem[{{Worthey}(1994)}]{Worthey94}
{Worthey}, G. 1994, \apjs, 95, 107

\bibitem[{{Worthey} {et~al.}(1994){Worthey}, {Faber}, {Gonzalez}, \&
  {Burstein}}]{Worthey94b}
{Worthey}, G., {Faber}, S.~M., {Gonzalez}, J.~J., \& {Burstein}, D. 1994,
  \apjs, 94, 687

\bibitem[{{Worthey} {et~al.}(2013){Worthey}, {Tang}, \& {Serven}}]{Worthey13}
{Worthey}, G., {Tang}, B., \& {Serven}, J. 2013, ArXiv:1303.2603

\bibitem[{{York} {et~al.}(2000)}]{York00}
{York}, D.~G. {et~al.} 2000, \aj, 120, 1579

\bibitem[{{Zoccali} {et~al.}(2004){Zoccali}, {Barbuy}, {Hill}, {Ortolani},
  {Renzini}, {Bica}, {Momany}, {Pasquini}, {Minniti}, \& {Rich}}]{Zoccali04}
{Zoccali}, M., {Barbuy}, B., {Hill}, V., {Ortolani}, S., {Renzini}, A., {Bica},
  E., {Momany}, Y., {Pasquini}, L., {Minniti}, D., \& {Rich}, R.~M. 2004, \aap,
  423, 507

\end{thebibliography}

\end{document}